\pdfoutput=1
\documentclass[reprint,nofootinbib]{revtex4-2}

\usepackage{graphicx}
\usepackage{dcolumn}
\usepackage{booktabs}
\usepackage{bm}
\usepackage{hyperref}
\usepackage{autobreak}
\usepackage{xcolor}
\usepackage{physics}
\usepackage{amsmath, amssymb} 
\usepackage{mathtools}
\usepackage{ragged2e}
\usepackage{soul} 
\usepackage{tikz,tkz-tab,amsmath}
\newcommand{\ds}{\displaystyle}
\usepackage{tikz,lipsum,lmodern}
\usepackage[T1]{fontenc}
\usepackage[most]{tcolorbox}
\usepackage{lmodern} 
\usepackage{float}
\usepackage{chemfig}  
\usepackage{wrapfig}
\usepackage{footmisc}

\tcbset{colback=green!5,colframe=green!35!black,fonttitle=\bfseries}

\NewTColorBox{NewBoxacc}{ s O{!htbp} }{%
  floatplacement={#2},
  IfBooleanTF={#1}{float*,width=\textwidth}{float}, colback=yellow!5!white,colframe=yellow!75!black,enhanced jigsaw, breakable, pad at break*=1mm,title=\centerline{\textbf{Box 1: Efficient Computation of Weak-Interaction Rates}} \label{app:accuracy_comput}
  }

\NewTColorBox{Reactions}{ s O{!htbp} }{%
  floatplacement={#2},
  IfBooleanTF={#1}{float*,width=\textwidth}{float}, colback=pink!5!white,colframe=pink!80!black,title=\centerline{\textbf{Box 2: Nuclear Reactions and Rates}} \label{app:reactions}
  }

\NewTColorBox{Decoupling}{s O {!htbp} }{
  floatplacement={#2},
  IfBooleanTF={#1}{float*,width=\textwidth}{float}, colback=green!5,colframe=green!35!black,enhanced jigsaw, breakable, pad at break*=1mm,title=\section{Neutrino Decoupling Thermodynamics} 
  }

\NewTColorBox{StiffCombined}{ s O{!htbp} }{%
  floatplacement={#2},
  IfBooleanTF={#1}{float*,width=\textwidth}{float}, colback=blue!5,colframe=blue!35!black,enhanced jigsaw, breakable, pad at break*=1mm,title=\centerline{\textbf{Box 3: Solving Stiff Differential Equations}} \label{app:stiff-combined}
  }

\def\eV{\, {\rm eV}}
\def\MeV{\,{\rm MeV}}
\def\GeV{\,{\rm GeV}}

\def\sec{\,{\rm s}}
\def\K{{\,\rm K}}
\def\mpl{{m_{\rm Pl}}}
\def\Neff{N_{\rm eff}}

\newcommand{\kmsMpc}{\,{\rm km\ s^{-1}Mpc^{-1}}}

\begin{document}

\preprint{000-000-000}

\title{\texttt{BBN-simple}:  How to Bake a Universe-Sized Cake}


\author{Aidan Meador-Woodruff}
 \email{ameadorw@ur.rochester.edu}
\affiliation{
Department of Physics and Leinweber Center for Theoretical Physics, University of Michigan, 450 Church St, Ann Arbor, MI 48109
}
\affiliation{Department of Physics and Astronomy, University of Rochester, 500 Wilson Blvd, Rochester, NY 14611}
\author{Dragan Huterer}%
 \email{huterer@umich.edu}
\affiliation{
Department of Physics and Leinweber Center for Theoretical Physics, University of Michigan, 450 Church St, Ann Arbor, MI 48109
}

\date{\today}

\begin{abstract}
Big Bang Nucleosynthesis (BBN), the process of creation of lightest elements in the early universe, is a highly robust, precise, and ultimately successful theory that forms one of the three pillars of the standard hot-Big-Bang cosmological model. Existing theoretical treatments of BBN and the associated computer codes are accurate and flexible, but are typically highly technical and opaque, and not suitable for pedagogical understanding of the BBN. Here we present \texttt{BBN-simple} -- a from-scratch numerical calculation of the lightest element abundances  pitched at an advanced undergraduate or beginning graduate level. We review the physics of the early universe relevant for BBN, provide information about the reaction rates, and discuss computational-mathematics background that is essential in setting up a BBN calculation. We calculate the abundances of the principal nuclear species in a standard cosmological model, and find a reasonably good agreement with public precision-level  BBN codes. A condensed version of this paper and associated snippets of computer code are given at \url{http://www-personal.umich.edu/~aidanmw/}.
\end{abstract}

\maketitle

The question of when the elements of the periodic table formed is both basic
and profound, and has been on scientists' mind for a long time. The correct
answer only arrived around the mid-20th century. \textit{Most} of the familiar
elements --- those up to iron, the most stable element --- formed in stars, in
nuclear reactions that take place for up to millions of years.  Stars burn
hydrogen to helium (which the Sun is doing busily at the moment); sufficiently
massive stars then burn helium to carbon and oxygen, et cetera, all the way up
to iron. However, the lightest elements in the universe - hydrogen, helium, as well as smaller amounts of a few more isotopes (from deuterium up to beryllium) were formed in the process of Big Bang Nucleosynthesis (BBN), occurring within the window of 1 second to 20 minutes after the Big Bang. 


The foundation for BBN calculations was set by the legendary $\alpha\beta\gamma$ paper \cite{Alpher:1948ve} which laid out the basics of element formation in the early universe. The physics of the BBN was refined and better understood in numerical treatments  spearheaded by George Gamow's group \cite{Gamow:1948pob,Alpher:1948srz,Alpher:1948yqy,Hayashi:1950lqo,Alpher:1953zz}, but with the physical reasoning that was still not wholly correct \cite{Turner:2021roa}. The BBN theory took its essentially modern form in the paper by Wagoner, Hoyle, and Fowler \cite{Wagoner67}, 
and by work of Jim Peebles \cite{Peebles:1966zz}. 
The key input to BBN are the nuclear reaction rates which, between the 1960s and 80s, were found to a (reasonably) high precision by Fowler et al.\ in the series of five papers 
\cite{FCZI, FCZII, FCZIII, FCZIV, FCZV}. These developments, along with the establishment of the standard cosmological model of the early universe, led to the theory of BBN in its present form. This theoretical background, along with comparison to observations, has been summarized in a suite of excellent BBN review papers (e.g.\ \cite{Copi:1994ev,Olive:1999ij,Walker:1991ap,Cyburt:2015mya}) and textbooks (e.g.\ \cite{Kolb:1990vq}). Precision-level BBN calculations (e.g.\ \cite{Lopez:1998vk,Burles:1999zt,Burles:2000ju,Burles:2000zk}), combined with an improved assessment of nuclear-reaction rates \cite{Adelberger:2010qa} and accurate measurements of the abundances of light elements  \cite{Pettini:2001yu,Fumagalli:2011iw,Noterdaeme:2012pa,Izotov:2014fga,Aver:2015iza,Cooke:2017cwo,Mossa:2020gjc} have enabled very accurate predictions for the physical quantities in the standard model of cosmology \cite{Nollett:2011aa,Schoneberg:2024ifp}, and can further be used to probe new physics \cite{Iocco:2008va,Jedamzik:2009uy,Pospelov:2010hj}. Finally, specialized computer codes enable fast and accurate evaluation of BBN abundances; these include the Kawano code
\cite{Kawano:1992ua} which is an extension of Wagoner's original code, \texttt{alterbbn} \cite{Arbey:2011nf,Arbey:2018zfh}, \texttt{PArthENoPE} \cite{Pisanti:2007hk},  and \texttt{PRIMAT} \cite{Pitrou:2018cgg}.

The basic blocks of the BBN calculations involve nuclear reactions governed by  laws of early-universe thermodynamics. Due mainly to the very low baryon-to-photon ratio (about two billion photons for each baryon), as well as the combination of a rapidly falling temperature and the consequently increased Coulomb barrier for creating heavier nuclei, BBN is inefficient and results in non-negligible abundances for only a handful of the lightest elements in the universe. The end result of the BBN calculation are quantitative predictions for the abundances of lightest elements, often shown in a classic plot of the abundances as a function of baryon density, $\rho_b$. Measurements of the abundances of these elements can then be used to constrain the expansion rate $H$ during the BBN era. The measured abundances also constrain the physical baryon density today, $\rho_b\propto \Omega_bh^2$, where $\Omega_b$ is the present-day baryon density relative to critical and $h$ is the Hubble constant in units of $100\kmsMpc$.   The BBN constraint on $\Omega_bh^2$ can then be used as a very powerful prior on other cosmological measurements. The importance of the BBN determination of baryon density has become magnified recently, with the realization that a prior on $\Omega_bh^2$ enables a measurement of the Hubble constant that is independent of either that from the cosmic microwave background anisotropies or distance ladder measurements \cite{DES:2017txv,DESI:2024mwx}, and can thus help weigh in on the discrepancy between these two kinds of measurements (the "Hubble tension" \cite{DiValentino:2021izs}).  
Precision measurements of the abundance of lightest elements, combined with BBN theory, have therefore essentially fixed one of the key parameters of the standard cosmological model. 

While BBN is therefore a mature and extremely successful cosmological probe, one remaining challenge is paradoxically pedagogical.  
The background theory, and especially the numerical calculation, are technical and difficult to adequately explain to students of cosmology in full detail. Theoretical treatments in textbooks \cite{Kolb:1990vq,Mukhanov:2005sc} and the ``no computer'' calculations \cite{Esmailzadeh:1990hf,Mukhanov:2003xs} do produce reasonably accurate results, but seem too technical and equation-heavy to be reasonably covered in a graduate course or a homework assignment. Other treatments (say \cite{Ryden,Huterer:2023mmv}) cover the matter pedagogically, but without sufficient detail required to actually perform a BBN calculation. Moreover, the  differential equations that governs the temporal evolution of abundances are ``stiff'', meaning that they include vastly different temperature scales. Solving these equations is challenging for a novice, at least when using standard numerical methods (e.g.\ default built-in functions in Python's \texttt{numpy} or \texttt{scipy}); well-known integrators that students may have learned, such as Runge-Kutta 23 or 45, fail quickly (without unreasonably short time steps) due to this stiffness.

This is where the present paper comes in. We attempt to bridge the gap between the fundamentally technical nature of BBN physics and mathematics, and the desire to explain the material in a simple and easy-to-follow way. We wish to only assume the knowledge of standard intro-graduate-level cosmology, as well as not assume sophistication with solving intermediate differential equations. Yet we wish to produce a fully quantitative calculation of BBN abundances. 


The paper is organized as follows. In Sec.~\ref{sec:outline}, we present an outline of our approach. In Sec.~\ref{sec:thermo} we lay out the ingredients of early-universe thermodynamics which contain the physics and provide the necessary basic equations. In Sec.~\ref{sec:reactions} we describe the strong, electromagnetic, and weak nuclear reactions, and the formalism to keep track of them.  In Sec.~\ref{sec:results}, we present the results of our basic calculation, and make comparisons with professional BBN codes. We conclude in Sec.~\ref{sec:concl}.

\section{Outline of approach}
\label{sec:outline} 

Perhaps the most intuitive way to view BBN and its foundation is seeing it like baking a cake -- one with a lot of fermions and rapid thermonuclear reactions. A cake comes with a list of ingredients and a recipe of steps to make it. Here we provide an overview of these ingredients, and in subsequent sections provide the necessary details.

The basic ingredients for the BBN calculation include:

\begin{itemize}
\item \textbf{Thermodynamics:} Early-universe thermodynamics is the underlying physical foundation that provides the necessary key equations for a BBN calculation. In our cake analogy, it is (quite literally) preheating the oven and setting the timer. Thermodynamics is of crucial importance, as the reaction equations for the elements depend only on ambient temperature and density (or, more generally, temperature of different relativistic species), so tracking the evolution of temperature(s) with time is essential.
\item
\textbf{Initial Conditions:} Because the evolution of the abundances of nuclear species is governed by differential equations, it is clear that initial abundances are required for the BBN calculation. These initial abundances are given by the nuclear statistical equilibrium, and they obey Saha-like equations. We will return to these concepts shortly.
\item
\textbf{Weak, Strong, and Electromagnetic Rates:} Interaction rates are a crucial input in the nuclear reaction network. These rates are experimentally determined by measurement of reaction cross sections at various energies, which are then interpolated as a function of energy.

\item
\textbf{Numerical Methods:} Lastly, the process of integrating the network of coupled differential equations that govern the abundance of nuclear species is nontrivial due to different temperature scales that characterize the different reactions. Therefore, implicit differentiation will be required as opposed to more standard explicit ODE routines, which fail quickly for ``stiff'' equations. The reason for the stiffness can ultimately be traced to the fact that we are working around equilibrium, with positive and negative terms that   almost --- but not precisely --- balance.
\end{itemize}
We will be discussing each of these ingredients at some length in Secs.~\ref{sec:thermo} and \ref{sec:reactions}.

In our calculations, we have adopted the use of natural units, setting $c=\hbar=k_B=1$, where $c$ is the speed of light, $\hbar$ is the reduced Planck constant, and $k_B$ is the Boltzmann constant (for more on natural units, see e.g.\ Chapter 1 and Appendix A of \cite{Huterer:2023mmv}). We restore dimensionful constants where they are illustrative or where we believe confusion might arise. For the numerical calculations, where appropriate we have adopted a flat $\Lambda$CDM universe with concordance values of the cosmological parameters (notably, the present-day baryon number density), which we quote where appropriate.

\section{Thermodynamics}
\label{sec:thermo} 

To begin, we must lay the groundwork of thermodynamics in the early universe. In our cake analogy, this is the equivalent of creating our oven and preheating it to the appropriate temperature. The evolution of photon, neutrino, electron, and positron temperatures with respect to time are driven by the expansion of space after the Big Bang. The expansion, in turn, is governed by the Friedmann equations which relate the Hubble parameter $H$ and the energy density of mass/energy components in the universe that are important at that time. Historically, the foundational thermodynamic equations relating temperature and energy density in the context of expanding space were laid out by more than 70 years ago, see for example \cite{Alpher:1953zz}. We now summarize those results.

\subsection{Basics: the time-temperature relation}
\label{sec:t-T_relation}

The first ingredient that we will need is the relation between time $t$ and thermodynamic temperature of the universe $T$. To establish that relation, we will make use of Friedmann's equations which describe how matter and energy components in the universe affect its expansion rate. We adopt the standard notation with $a$ the scale factor\footnote{In historical literature, $a$ was denoted $R$ or, even earlier, $\ell$, such as in \citet{Alpher:1953zz}.}, and $H\equiv \dot{a}/a$ to be the Hubble parameter; both of them are a function of time $t$. Then the two Friedmann's equations read

\begin{eqnarray}
\label{eq:FI}
    H^2 \equiv \left(\frac{\dot a}{a}\right)^2 &=& \frac{8\pi G}{3}\rho\\[0.2cm]
    \frac{\ddot a}{a} &=& - \frac{4\pi G}{3} (\rho + 3P), 
    \label{eq:FII}
\end{eqnarray}
where $\rho$ is the total energy density of all species, while $P$ is their total pressure. In the early universe, photons and relativistic neutrinos dominate the energy budget, and for both of them $P=\rho/3$.

In addition to the Friedmann equations, also useful is the continuity equation
\begin{equation}
\dot \rho + 3H(\rho + P) = 0
\label{eq:continuity}
\end{equation}
which however is not independent of the former two (any \textit{two} of the equations (\ref{eq:FI})-(\ref{eq:continuity}) are independent).

We now have all the tools to calculate $dT/dt$. Adopting the chain rule, we obtain
\begin{equation}
\begin{aligned}
    \label{eq:Tt-relation}
    \frac{dT}{dt} &= \frac{dT}{d\rho}\frac{d\rho}{dt}  = 3H(\rho + P)  \frac{dT}{d\rho} \\[0.2cm]
    &= (24 \pi G \rho)^{1/2} \left(\rho + P\right) \left( \frac{d\rho}{dT}\right)^{-1}. 
\end{aligned}
\end{equation}
In the era when $T\gg 1\MeV$, all of the relevant species (electrons and positrons, neutrinos, as well as photons) are relativistic ($T\gg m$) so that $P= \rho/3$. Moreover, for relativistic species the energy density goes as temperature to the fourth power
\begin{equation}
\rho = \sum_i \rho_i = \frac{\pi^2}{30} g_{*} T^4,
\label{eq:rho_rel}
\end{equation}
where $g_*$ is the effective number of relativistic degrees of freedom. The parameter $g_*$ changes over time, and decreases when the temperature falls below the mass of each particle. Just before the BBN (at $T\gg 1\MeV$), $g_*$ is equal to 10.75 and steadily falls around the time of electron-positron annihilation. Following prior work \cite{Wagoner67,Kawano:1992ua} and for simplicity, we assume a fixed value $g_*\simeq 9$.\footnote{The effective number of relativistic species $g_*$ clearly depends on the number of neutrino species as these particles are \textit{very} relativistic during BBN. In the standard model of particle physics, there are three neutrino species, but due to subtle quantum effects the number that enters $g_*$ evaluates to a non-integer, $\Neff=3.046$. If there are new particles that decay into photons or additional relativistic species in their own right, then $\Neff$ can be different from the standard value. Therefore, measurements of the abundances of light elements compared to BBN theory (which is sensitive to the expansion rate $H$ and hence $g_*$ and $\Neff$) provide important information in constraining this parameter. For our key results summarized in Sec.~\ref{sec:results}, we assume the standard value $\Neff=3.046$. \label{foot:Neff}}

From Eqs.~(\ref{eq:Tt-relation}) and (\ref{eq:rho_rel}), the temperature-time relation can be easily integrated to get
\begin{equation}
T^2 = \sqrt{\frac{45}{16 \pi^3 G g_*}}\,t^{-1}. 
\label{eq:tT-Rel} 
\end{equation}
Note that $G^{-1/2}\equiv \mpl = 1.22\times 10^{19}\GeV$ in natural units with $c=\hbar=1$. Moreover, expressing temperature in units of $10^9\K$ and time in seconds, and using the conversions $1 \sec=1.519\times 10^{15}\eV$ and $1 \K = 8.619 \times 10^{-5}\eV$, we get

\begin{equation}
   \left(\frac{T}{10^9 \K}\right)^2 \simeq 325.4\, g_*^{-1/2} \left(\frac{t}{1 \sec}\right )^{-1}
\end{equation}
or, equivalently, 
\begin{equation}
  \left (\frac{T}{\MeV}\right )^{2} \simeq 2.42\,g_*^{-1/2}\left (\frac{t}{1\sec}\right )^{-1},
  \label{eq:t_T_relation}
\end{equation}
where, recall, $g_*\simeq 9$. 

\subsection{Neutrino Decoupling}
\label{sec:nu_decoupling} 

At high temperature ($T\gg 1\MeV$) and early time ($t\ll 1\sec$), weak interactions keep neutrinos in equilibrium with the thermal bath. However, as the universe expands and cools, the weak-interaction rate $\Gamma(t)$ falls off faster than the expansion rate $H(t)$. When $\Gamma<H$, at a temperature of about an MeV (or $10^{10}\K$), neutrinos fall out of equilibrium, and their temperature thereafter evolves differently from that of the photons. This so-called neutrino decoupling occurs during the early stages of BBN (see e.g.\ Chapter 5 of \cite{Huterer:2023mmv} or, for a more detailed treatment, Chapter 5 of \cite{Kolb:1990vq}). 

Our goal is to establish the relation between the neutrino temperature $T_\nu$ and photon temperature $T\equiv T_\gamma$ as a function of time. We need this because both of these temperatures are required as an input in weak  interaction rates that we describe in Sec.~\ref{sec:reactions}.

It turns out that the relation between the neutrino and photon temperatures depends on the energy densities and pressures of the dominant components in the universe. As shown in \citet{Wagoner67}, the relation between the photon and neutrino temperature is

\begin{equation}
\frac{T_\gamma}{T_\nu} = \left(\frac{11}{4} \frac{\rho_\gamma + P_\gamma }{\rho_{\rm rad} + P_{\rm rad} }\right)^{1/3},
\label{eq:PhotonNeutrinoRelationship}
\end{equation}
where $\rho_{\rm rad} = \rho_\gamma + \rho_{e^+} + \rho_{e^-}$ is the total radiation energy density, and similar for $P_{\rm rad}$. 


Asymptotic values of the ratio in Eq.~(\ref{eq:PhotonNeutrinoRelationship}) are well known: for $T\gg 1\MeV$, the neutrinos are in equilibrium with the cosmic radiation fluid and $T_\gamma/T_\nu=1$. And, at temperatures  below an MeV, electrons and positrons annihilate (since $m_{e^-}=m_{e^+}\simeq 0.5\MeV$) and give their energy to the photons, but (to a good approximation) not to the neutrinos which had just decoupled.  
For the electron-positron-photon bath, a
a simple accounting of the relativistic degrees of freedom (e.g.\ \cite{Huterer:2023mmv}) gives
\begin{equation}
  g_{*S}^{\rm th} = 
\left \{  
\begin{array}{cl}
   2+\ds\frac{7}{8}\times2\times 2=\ds\frac{11}{2}  & (T\gtrsim  m_e) \\[0.25cm]
   2  & (T\lesssim m_e),
\end{array} 
\right .
\label{eq:gstar_ee_dec}
\end{equation}
where, in the latter case, only two polarizations of photons contribute, while in the former case the factors of two are also accounting for the electrons and positrons, each with two spin states. Because the entropy $S=g_{*S}^{\rm th} (aT_\gamma)^3$ is conserved, a step down in $g_{*S}^{\rm th}$ implies a step up in the photon temperature, so that  $T_\gamma/T_\nu = (g_{*S, \,T\gtrsim  m_e}^{\rm th}/g_{*S, \, T\lesssim  m_e}^{\rm th})^{1/3}=(11/4)^{1/3}$; see Fig.~\ref{fig:decoupling}.

A full numerical, time-dependent relation for the $T_\gamma/T_\nu$ ratio can be obtained as follows. We need to evaluate the expression in Eq.~(\ref{eq:PhotonNeutrinoRelationship}), for which we need to know the energy densities and pressures of photons, electrons/positrons, and neutrinos. For photons, the expression for the energy density is the familiar 
\begin{equation}
    \rho_\gamma = \frac{\pi^2}{15} T_\gamma^4
\end{equation}
which is the expression in Eq.~(\ref{eq:rho_rel}) with two relativistic degrees of freedom ($g=2$ for photons because they have two polarizations). This can be further cast in units of $T_9$ (temperature in units of $10^9\K$), where it becomes $\rho_\gamma \simeq 8.42\,T_9^4$. Similarly, the photon pressure and energy density obey the familiar relationship
\begin{equation}
    P_\gamma = \frac{1}{3} \rho_\gamma.
\end{equation}
For non-degenerate neutrino species, the expression for their energy density is similarly given by
\begin{equation}
    \rho_{\nu} = \frac{7}{8} \frac{\pi^2}{15} N_{\rm eff} T_\nu^4
\end{equation}
where $g=7/8$ is the effective number of relativistic degrees of freedom, and $N_{\rm eff}$ is the number of neutrino species. We take to be $N_{\rm eff} = 3.046$.

\begin{NewBoxacc}*
\textbf{Background.} Equation~(\ref{eq:weak_rxns}) lists the weak interactions between the protons and neutrons. The rates for these reactions are given in Eqs.~(\ref{eq:weak_rates})-(\ref{eq:rev_weak}) and explained in Section (\ref{sec:weakrxns}), but these integrals are somewhat cumbersome to evaluate efficiently at every value of the photon and neutrino temperature. \\

\textbf{Kawano approximation.} To speed up the computation, an optional approximation of these rates was offered in \citet{Kawano:1992ua}: 
\begin{equation*}
\begin{aligned}
    \lambda_{n\rightarrow p} &= \frac{1}{\tau} \left(1 + \frac{0.565}{z} - \frac{6.382}{z^2} + \frac{11.108}{z^3} + \frac{36.492}{z^4} + \frac{27.512}{z^5} \right) \\[0.2cm]
    \lambda_{p\rightarrow n} &= \frac{1}{\tau}\left( \frac{5.252}{z} - \frac{16.229}{z^2}+\frac{18.059}{z^3} +\frac{34.181}{z^4} +\frac{27.617}{z^5}\right)e^{-qz}, 
    \end{aligned}
    \label{eq:approx_weak_rates}
\end{equation*}
where $q = (m_n - m_p)/m_e$ is the neutron-proton mass difference in units of the electron mass, $z = m_e/T_\gamma$ is the ratio of electron mass and photon temperature, and $\tau$ is the lifetime of the neutron. Replacing an integral with the polynomial above vastly improves the computational speed of these rates, and leads to only fractions of a percentage loss of accuracy in the final abundances. \\

\textbf{Evaluation with Gaussian quadratures.}
Here, however, we employ a comparably efficient method with a higher accuracy than the analytic expressions above. We express the integrals in Eqs.~(\ref{eq:weak_rates})-(\ref{eq:rev_weak}) as using Gaussian quadratures, which approximate each integral as a sum. Specifically, the integral of some function $f$ over some interval can be written as
\begin{equation}
    \label{eq:quadrature}
    \int_{a}^b f(x) ~ \dd x \approx \sum_{i=1}^N w_i f(x_i),
\end{equation}
where $w_i$ are the Gaussian-quadrature weights, and $x_i$ are the roots of the $N$-th degree Legendre polynomial. The choice of $N$ is arbitrary, with a higher $N$ being more accurate at the cost of computational speed. Even for large numbers of weights, this method will still be much faster than solving the integral analytically. This method is also offered in Kawano's code. \\ 



\begin{wrapfigure}{l}{6.5cm}
\includegraphics[width=6.5cm]{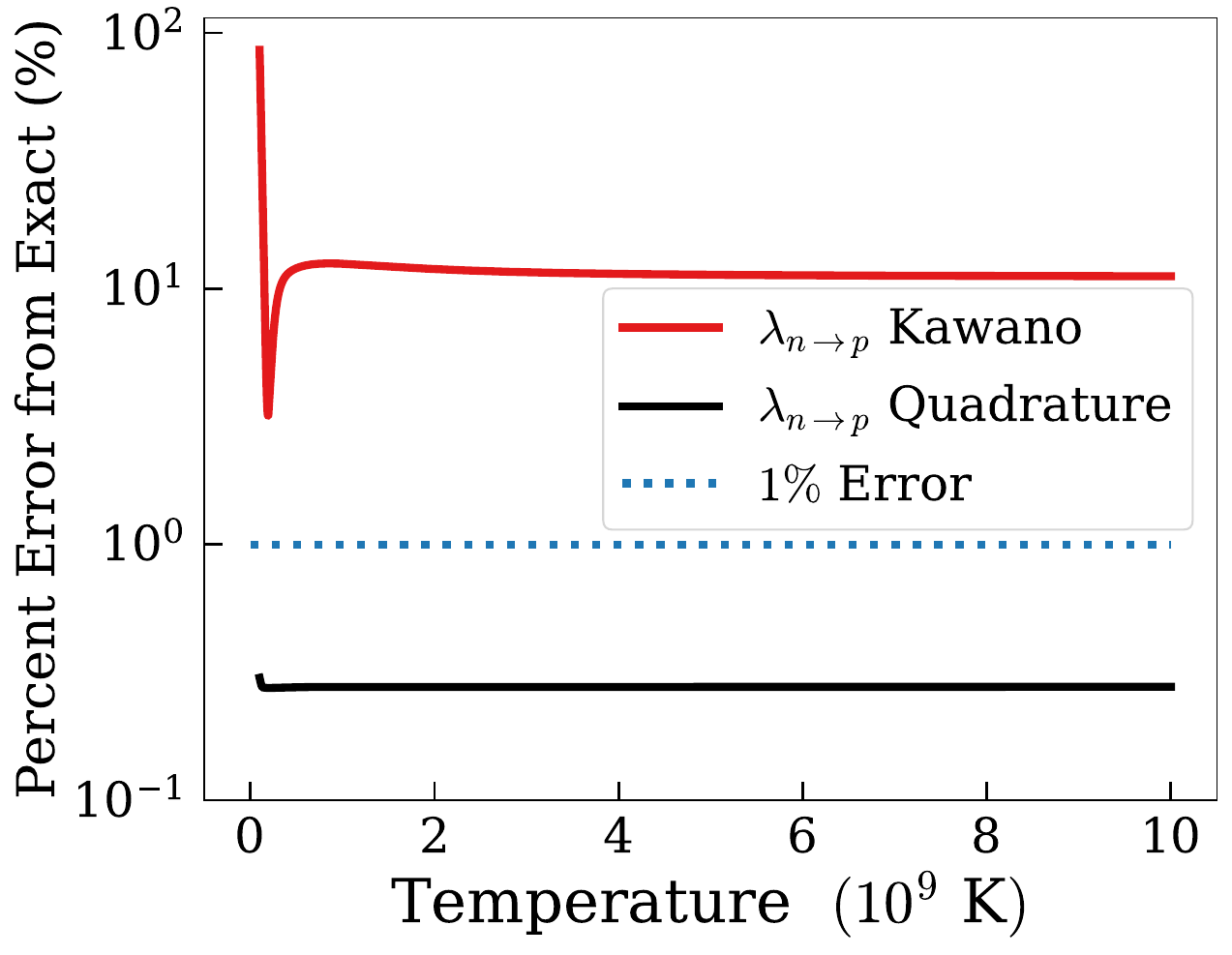}
\end{wrapfigure}

\textbf{Accuracy.}
The plot on the left shows the accuracy of the calculated cross-section for conversion of neutrons to protons ($\lambda_{n \rightarrow p}$) as a function of temperature around the time of the BBN. The accuracy of the Kawano approximation is only about 10\% and depends on $T$, while that of our quadrature calculation stays constant at about 0.5\%. Our numerical evaluation approximation (with $N=64$ quadrature points), while slower than the analytic expressions above, remains quite fast (about 2.7 ms for each temperature step). 

We find that the final element abundances (shown in Fig.~\ref{fig:Abundances}) are not strongly affected by accuracy of the weak-interaction rates. Nevertheless, we find that the inaccuracy in the Kawano approximation leads the final abundances that are about 0.5\% different from those using the quadrature to compute the rate integrals. Therefore, there is no reason not to use the latter methodology. 



\end{NewBoxacc}

Electrons and positrons start out relativistic, but we cannot ignore their rest masses. For each of these species, 
\begin{equation}
\rho =
    \frac{g}{(2\pi)^3}\int_0^\infty 
    \frac{E(p)}{e^{E(p)/T} + 1}\,d^3p,
\end{equation}
where $p$ is momentum, related to energy by the familiar relativistic relation $E=\sqrt{p^2+m^2}$. [To convert this expression and those below to MKS units, we would multiply each power of $T$ by $(\hbar c)/k_B$.] Further,  the pressure is given by
\begin{equation}
    P = \frac{g}{(2\pi)^3} \int f(p) \frac{p^2}{3 E(p)} \, d^3 p,
\end{equation}
where $f$ is the phase-space distribution function. For particles that exchange energy and momentum efficiently, the distribution function is
\begin{equation}
  f(p) = \frac{1}{e^{E(p)/T}+ 1},
\end{equation}
where the plus sign in the denominator indicates that electrons and positrons are fermions. Introducing the substitutions
\begin{equation}
    x\equiv \frac{m}{T};\qquad y\equiv\frac{p}{T}
\end{equation}
and adopting $g=4$ (two spin states for both electrons and positrons), the sum of the electron and positron energy densities can be rewritten as 
\begin{equation}
    \rho_{e^-} + \rho_{e^+} = \frac{2}{\pi^2} T^4_{\gamma}\int_0^\infty \frac{y^2 \sqrt{x^2 + y^2}}{\exp (\sqrt{x^2 + y^2}) + 1}dy
    \label{eq:rho_elec_positr}
\end{equation}
while their combined pressure becomes
\begin{equation}
    P_{e^- + e^+} = \frac{2}{3 \pi^2 } T_\gamma^4  \int_0^\infty  \!\! \frac{y^2 ~  dy}{\sqrt{x^2 + y^2} \left(\exp(\sqrt{x^2 + y^2}) + 1\right)}.
    \label{eq:P_elec_positr}
\end{equation}







The integrals in Eq.~(\ref{eq:rho_elec_positr}) and \ref{eq:P_elec_positr} need to be evaluated at each temperature/time step, so speeding up these evaluations is useful. An early such treatment was discussed by \citet{FowlerHoyle1964} in their Appendix B, as well as \citet{Kawano:1992ua} who made use of the modified Bessel functions\footnote{In fact, expansions of the Fermi-Dirac integrals in terms of modified Bessel functions date back to the 1930s \cite{1942psd..book.....C}.}. Another approach would be to use Gaussian quadrature \footnote{More precisely, we used Gauss-Laguerre quadrature to handle the limits of the integral.}, which is what we adopted in our code.


\begin{figure}[t]
    \centering
    \includegraphics[width=\linewidth]{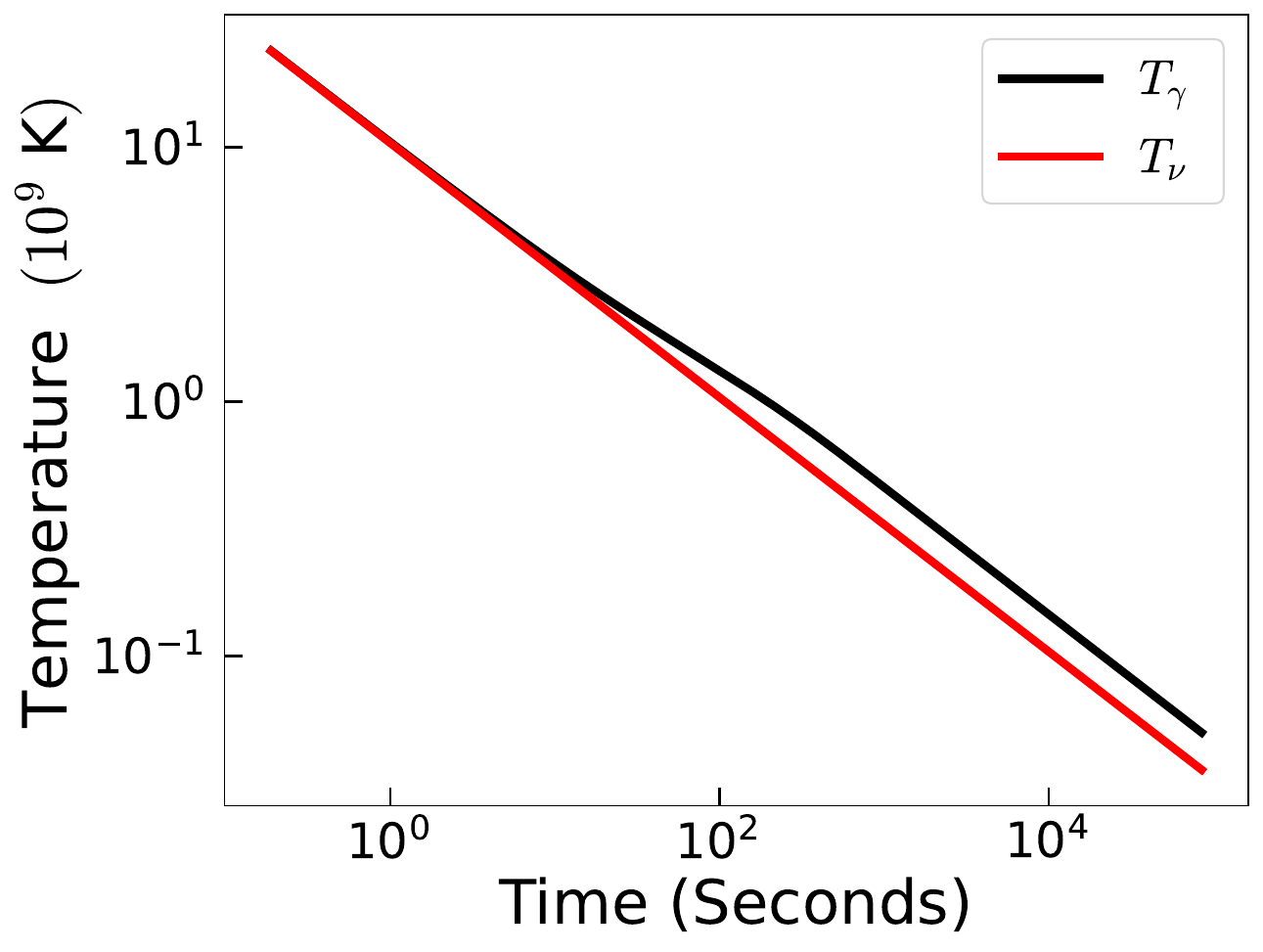}
    \caption{Neutrino and photon temperatures as a function of time. The neutrino decoupling happens at $t \simeq 1 - 10 ~ \rm{s}$ after the Big Bang. The late-time ratio of photon and neutrino temperatures is $(11/4)^{1/3}$.}
    \label{fig:decoupling}
\end{figure}

With the expressions for the electron/positron and photon density and pressure as a function of temperature, Eq.~(\ref{eq:PhotonNeutrinoRelationship}) gives the desired expression for the neutrino temperature. Its evolution vs.\ the photon temperature is shown in Fig.~\ref{fig:decoupling}.

\section{Nuclear Reactions}
\label{sec:reactions} 

At temperatures greater than $10 \MeV$, rapid weak interactions keep protons and neutrons in nearly equal abundance. As the universe cools below an MeV, these particles fall out of thermal equilibrium, causing the neutrons to ``freeze-out'' when the temperature can no longer sustain the weak reactions, at around $0.8 ~\MeV$. [The abundance of free neutrons continues to fall slowly at this time due to their beta decay.]

As the universe sufficiently cools and weak interactions slow, protons and neutrons begin to fuse through strong and electromagnetic interactions, marking the onset of BBN. This phase occurs within a narrow time window due to the rapid cooling and expansion of the universe. By the time the temperature drops to around $30 ~ \rm{keV}$, the universe is no longer hot enough to  sustain a significant rate of nuclear reactions. 

Hydrogen (nucleus with just one proton) and helium-4 (two protons, two neutrons) are the most significant products of BBN owing to their high binding energy and stability. Deuterium and tritium, nuclei with one proton and (respectively) one and two neutrons, are also produced, though most of these nuclei quickly fuse into helium-4.  Lithium and beryllium form in small quantities (their BBN abundances are ballpark $10^{-10}$ relative to hydrogen), and all nuclei heavier than them form in trace amounts, as their formation is limited by the decreasing temperature.

The rapid cooling of the universe curtails the synthesis of elements much heavier than beryllium. These heavier elements are instead formed in later astrophysical processes, primarily stellar fusion. 

In this section we cover the basics of nuclear reactions that play important roles during BBN. This enables us to set up the formalism to track the abundance of elements in cosmic time during the time of BBN.

\subsection{Nuclear statistical equilibrium}
\label{sec:weakrxns} 

At high temperatures ($T\gg 1\MeV$) that exceed the $Q$ values (the amounts of energy absorbed or released) of nuclear reactions, all nuclei are in equilibrium that is governed by the ambient temperature as well as the binding energy and spin state of each species. In this so-called nuclear statistical equilibrium, the number density of nuclear species is governed by the Maxwell-Boltzmann distribution. Likewise, the derivatives of their number densities is zero, a fact that will prove important later. Consequently, we can express the mass fraction of elements at the epoch just before the BBN by knowing only the basic quantum-mechanical properties of each nuclear species.

The most consequential species in this limit are protons and neutrons, which we consider first. 
They start out in chemical equilibrium at high temperature. As the universe cools, the neutron-to-proton ratio falls out of equilibrium, as the reactions keeping them in equilibrium cannot keep up with the expansion of the universe. 

In equilibrium, the ratio of protons and neutrons is 
\begin{equation}
    \frac{N_p}{N_n} = e^{Q/T_\gamma},
\end{equation}
where $Q$ is the mass difference of the neutron and the proton, $Q\equiv m_n - m_p = 1.29\MeV$.


More generally, the abundance of all species in nuclear statistical equilibrium is governed by the Maxwell-Boltzmann distribution. For example, for deuterium -- nucleus with one proton and one neutron --  the Maxwell-Boltzmann expressions be evaluated to be (see e.g.\ Chapter 7 of \cite{Huterer:2023mmv})
\begin{equation}
\frac{N_D}{N_n N_p} = \frac{g_D}{g_n g_p} \left(\frac{2\pi m_D}{m_nm_p T}\right)^{3/2} e^{(m_n + m_p - m_D)/T}.
\label{eq:DMassFrac}
\end{equation}
In this expression:
\begin{itemize}
    \item  $N_n$, $N_p$, $N_D$ are the number densities of neutrons, protons, and deuterium respectively; 
    \item $m_n$, $m_p$, $m_D$ are the masses of neutrons, protons, and deuterium, and
    \item $T$ is the temperature in \MeV.
\end{itemize}

In equilibrium, it is often more convenient to express the number density of a species, $N_i$, with respect to the number density of all baryons, $N_b$. This expression works out to be (e.g. \cite{DodelsonSchmidt,Huterer:2023mmv}) 
\begin{equation}
    \frac{N_i}{N_b} \propto \eta_b^{{A_i}-1} \left(\frac{T}{m_n}\right)^{3/2} e^{B_i/T}. 
    \label{eq:nse}
\end{equation}
Here $\eta_b\equiv N_b/N_\gamma$ is the baryon-to-photon ratio, which we take to be $\eta_b\simeq 6.12\times 10^{-10}$ \cite{ParticleDataGroup:2020ssz}, 
$A_i$ is the mass number of species $i$, and $B_i$ is that species' binding energy. This relation is accurate when $\eta_b$ dominates over the exponential term, which is during the epoch preceding the BBN. Finally, note that $\eta_b$ and the baryon density can be traded off; specifically, $\Omega_b h^2\simeq 3.662\times 10^7\,\eta_b$ (see e.g.\ Eq.~(7.43) in \cite{Huterer:2023mmv}).

It is helpful to introduce some additional notation here: it is customary to denote the mass fraction of a baryonic species as its number density over the total sum of baryons. 
We extend the notation of the mass fraction from Eq. (\ref{eq:pn-mass}) for all species $i$ as
\begin{equation}
 X_i = A_i \frac{N_i}{N_b},
  \label{eq:massfrac}
\end{equation}
which must satisfy the condition
\begin{equation}
    \sum_i X_i = 1
\end{equation}
that enforces the contribution of all baryonic mass fractions be 100\% of the total. 
For elements of higher mass number than deuterium, the mass fraction expressions have many more terms than Eq.~(\ref{eq:DMassFrac}); the general expression for $X_i$ for any mass number is given in Chapter 4 of \citet{Kolb:1990vq}.

A few more words about conventions in the BBN field: an abundance of a species $i$ is often reported relative to that of hydrogen (that is, $X_i/X_p$). Moreover, it is also customary to denote the mass fraction of helium-4 as $Y_{\rm p} = X_{^4 {\rm He}}$, and not divide it by $X_p$.

\begin{Reactions}*
\textbf{Nuclear reactions.} Here we compile and list all twelve reactions for the associated principal elements that our simple BBN code contains. The rates are given in the table below, and the right column has references to papers where equations and expressions for these rates can be found. All the reactions between nuclei were taken from the ReacLib database \cite{Cyburt:2010}. This is essential input for any from-scratch BBN calculation.

\begin{table}[H]
\begin{ruledtabular}
\begin{tabular}{lcc}
\textrm{Number}&
\textrm{Reaction(s)} &\textrm{Source}\\
\colrule
1 &  Equation \eqref{eq:weak_rxns} & Appendix F of \citet{Kawano:1992ua} \\
2 & $p + n  \leftrightharpoons \rm{D} + \gamma $ &\citet{Cyburt:2010}\\
3 &  $\rm{D} + p \leftrightharpoons \mbox{$^3$He} + \gamma$ & \citet{Cyburt:2010}\\
4 &$\rm{D} + \rm{D} \leftrightharpoons n + \mbox{$^3$He}$ & \citet{Cyburt:2010}\\
5 & $\rm{D} + \rm{D} \leftrightharpoons p + \rm{T}$ & \citet{Cyburt:2010}\\
6 & $\rm{T} + \rm{D} \leftrightharpoons n + \mbox{$^4$He}$& \citet{Cyburt:2010}\\
7 & $\rm{T} + \mbox{$^4$He} \leftrightharpoons \mbox{$^7$Li} + \gamma$& \citet{Cyburt:2010}\\
8 & $^3\rm{He} + n \leftrightharpoons p + \rm{T}$& \citet{Cyburt:2010}\\
9 & $^3\rm{He} + \rm{D} \leftrightharpoons p + \mbox{$^4$He}$& \citet{Cyburt:2010}\\
10& $^3$He $+ \mbox{$^4$He} \leftrightharpoons \mbox{$^7$Be} + \gamma$ & \citet{Cyburt:2010}\\
11& $^7$Li $+ \rm{p} \leftrightharpoons \mbox{$^4$He} + \mbox{$^4$He}$& \citet{Cyburt:2010}\\
12& $^7 \rm{Be} + n \leftrightharpoons p + \mbox{$^7$Li}$&  \citet{Cyburt:2010}
\end{tabular}
\end{ruledtabular}
\end{table}

These reactions are also shown more graphically in the left panel just below. Note that the reaction number 1 comprises all weak-decay reactions, listed in Eq.~(\ref{eq:weak_rxns}). \\

\includegraphics[width=0.35\linewidth]{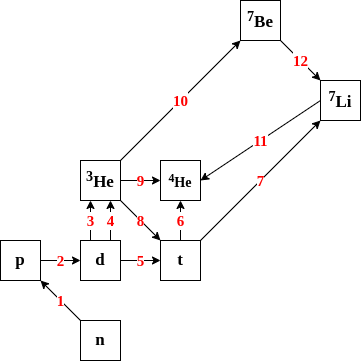} 
    \label{fig:BBNRxns}
\includegraphics[width=0.60\linewidth]{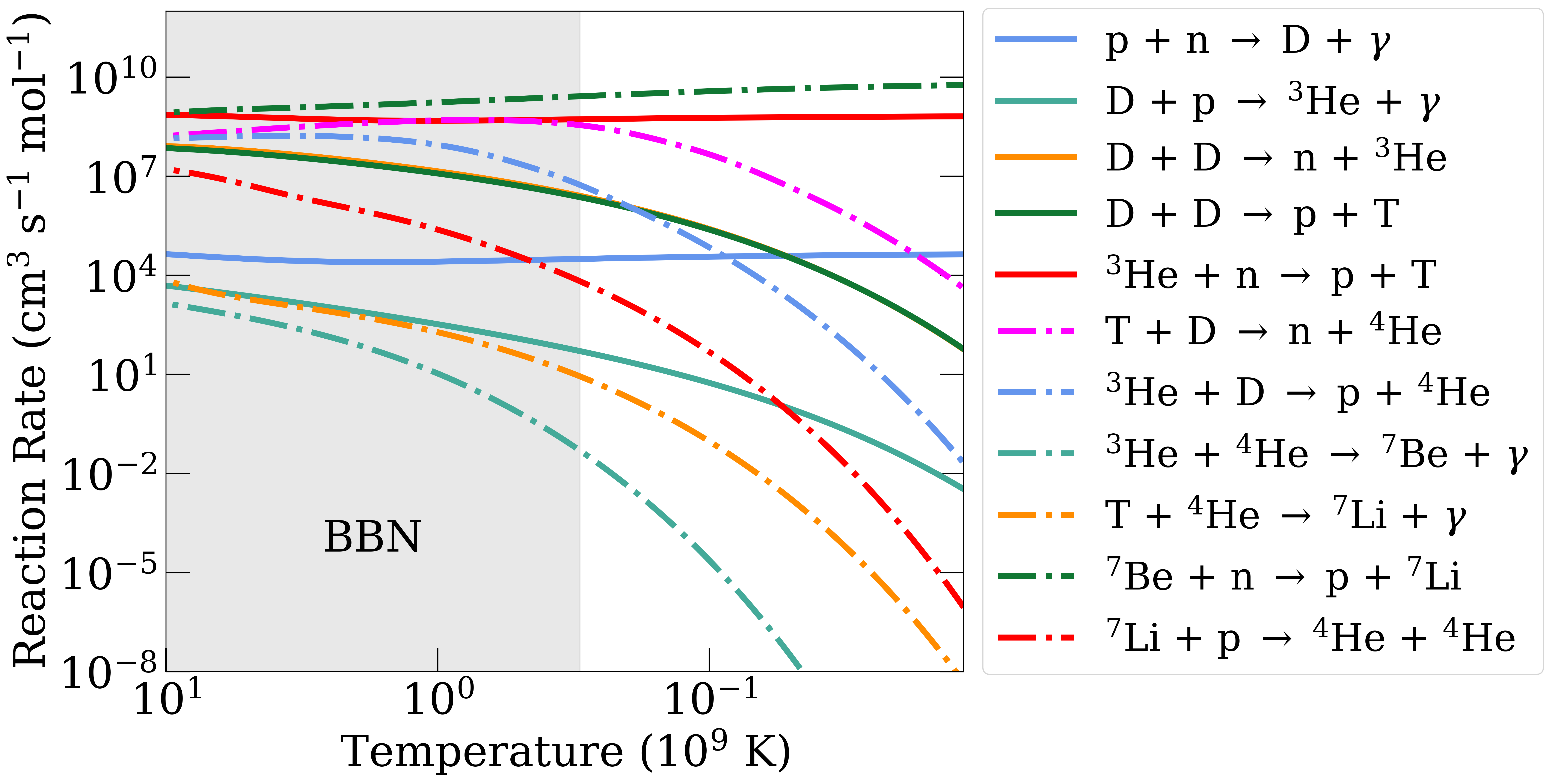}
    \label{fig:rates}
   
\vspace{0.4cm}
\textbf{Nuclear rates.}
The right panel above shows the principal reaction rates as a function of temperature around the time of the BBN. [This plot is a recreation of Figure 10 of \citet{SmithKawanoMalaney93}.] The rates are in units of cubic centimeters per second per mole ($\rm{cm}^3 \rm{s}^{-1} \rm{mol}^{-1}$). The reactions in the legend follow the same ordering as the enumerated reactions in the table above (and the chart on the left). 

\end{Reactions}

\subsection{Weak interactions and neutron decoupling}
\label{sec:weak_int} 

We next track the evolution of protons and neutrons as the temperature falls to become less than their mass difference $Q$. First, let us introduce the main quantities of interest, contributions to number density of protons and neutrons, with the general expressions defined in Eq.~(\ref{eq:massfrac}). At high temperature when protons and neutrons are the only nuclei of relevance, these evaluate to
\begin{equation}
\begin{aligned}    
    X_p &\equiv \frac{N_p}{N_n + N_p},\\[0.2cm]
    X_n &\equiv \frac{N_n}{N_n + N_p}.
\end{aligned}    
    \label{eq:pn-mass}
\end{equation}

There are six primary weak interactions between protons and neutrons (including forwards and reverse), namely
\begin{equation}
\begin{aligned}
\label{eq:weak_rxns}
n + \nu_e &\longleftrightarrow p + e^- \\
n + e^+ &\longleftrightarrow p + \overline{\nu}_e \\
n &\longleftrightarrow p + e^- + \overline{\nu}_e,
\end{aligned}
\end{equation}
where $e^-$ and $e^+$ refers to electrons and positrons, and $\nu_e$ and $\overline{\nu}$ to electron neutrinos and anti-neutrinos, respectively.

We denote the rate of conversion of neutrons into protons as $\lambda_{n \rightarrow p}$  and likewise $\lambda_{p \rightarrow n}$ as the rate of conversion of protons into neutrons. Each of these is the sum of their three respective reactions of the six primary reactions above. 
These rates can be analytically expressed, as described by 
Ref.~\cite{1983MNRAS.205..683S}; the rates are the integrals\footnote{It is worth noting that these rates are still an approximation neglecting nuclear recoil and QED corrections; the approximations lead to some inaccuracies, for example a 2\% underestimate in the abundance of $^4$He \cite{Lopez:1998vk}.}

\begin{equation}
\begin{split}
    \lambda_{p \rightarrow n} = K \int_1^\infty dx \frac{x(x+q)^2 (x^2 - 1)^{1/2}}{\left(1+e^{-xz}\right)[1 + e^{(x+q)z_{\nu} + \xi_e}]} \\[0.2cm]
    + K \int_1^\infty dx \frac{x(x-q)^2 (x^2 - 1)^{1/2}}{\left(1+e^{xz}\right)[1 + e^{-(x-q)z_{\nu} + \xi_e}]}
    \label{eq:weak_rates}
\end{split}
\end{equation}
and 

\begin{equation}
    \lambda_{n \rightarrow p} = \lambda_{p\rightarrow n}(-q,-\xi_e).
    \label{eq:rev_weak}
\end{equation}
Here we have introduced the following variables:
\begin{itemize}
    \item $q = (m_n - m_p)/m_e=Q/m_e$, the neutron-proton mass difference in units of the electron mass;
    \item $z = m_e/T_\gamma$ is the ratio of electron mass and photon temperature;
    \item $z_\nu = m_e/T_\nu$, which is the same as $z$ but for the neutrino temperature;
    \item $K$ is the normalization constant set so that these integrals asymptote at late (post-BBN) times to the inverse of the neutron decay lifetime, $1/\tau$; 
    \item $\xi_e$ is the chemical potential of the electrons. We have taken it to be zero, as relative to other contributions it is negligible.
\end{itemize}

Eqs.~(\ref{eq:weak_rates}) and (\ref{eq:rev_weak}) describe the rates for all six reactions in Eq.~(\ref{eq:weak_rxns}). It is somewhat cumbersome to evaluate the integrals in Eq.~(\ref{eq:rev_weak}) and Eq.~(\ref{eq:weak_rates}) at every value of the photon and neutrino temperature, so approximate and fast methods to do so have been devised. A comparison of the approximations for the reaction rates, and the effect on the numerical abundances of the lightest nuclei, are further discussed in Box 1. 

\begin{figure*}[t]
    \centering
    \includegraphics[width=0.45\linewidth]{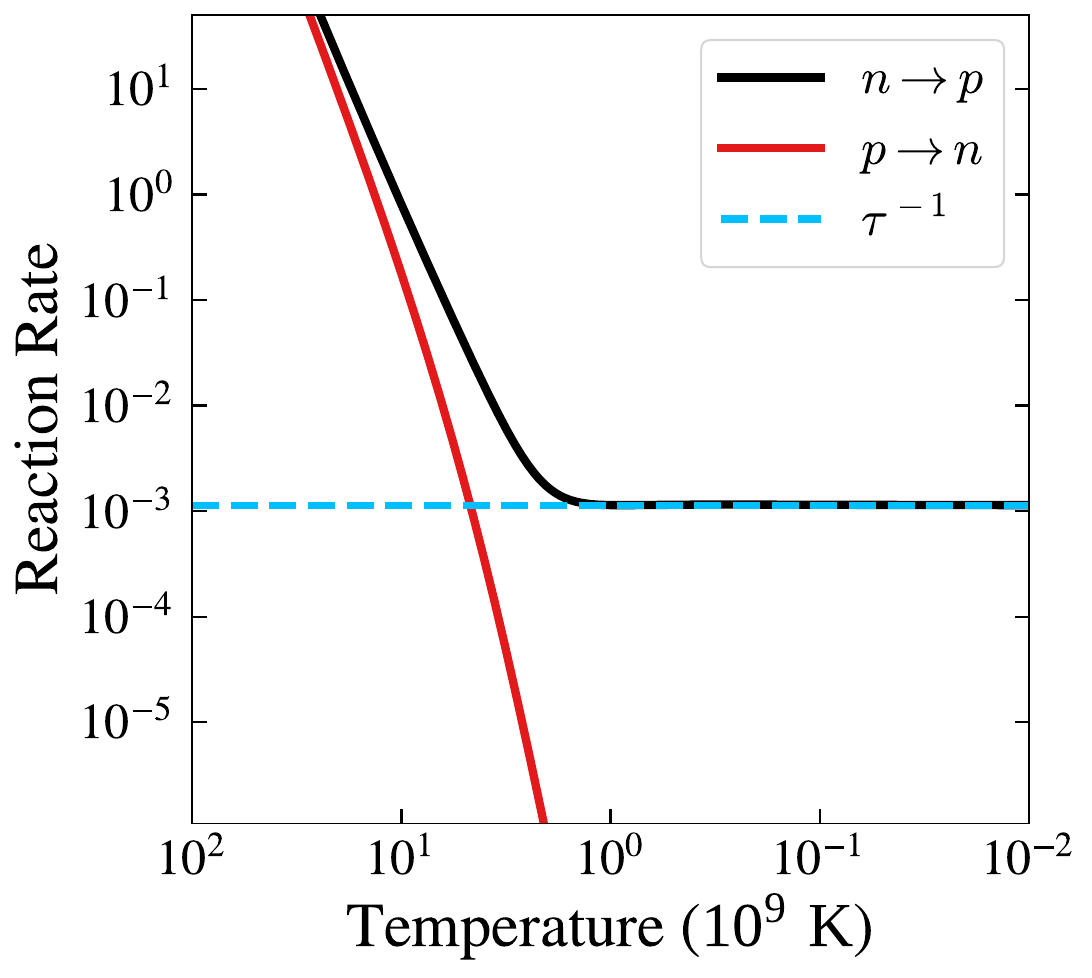}
    \includegraphics[width=0.5\linewidth]{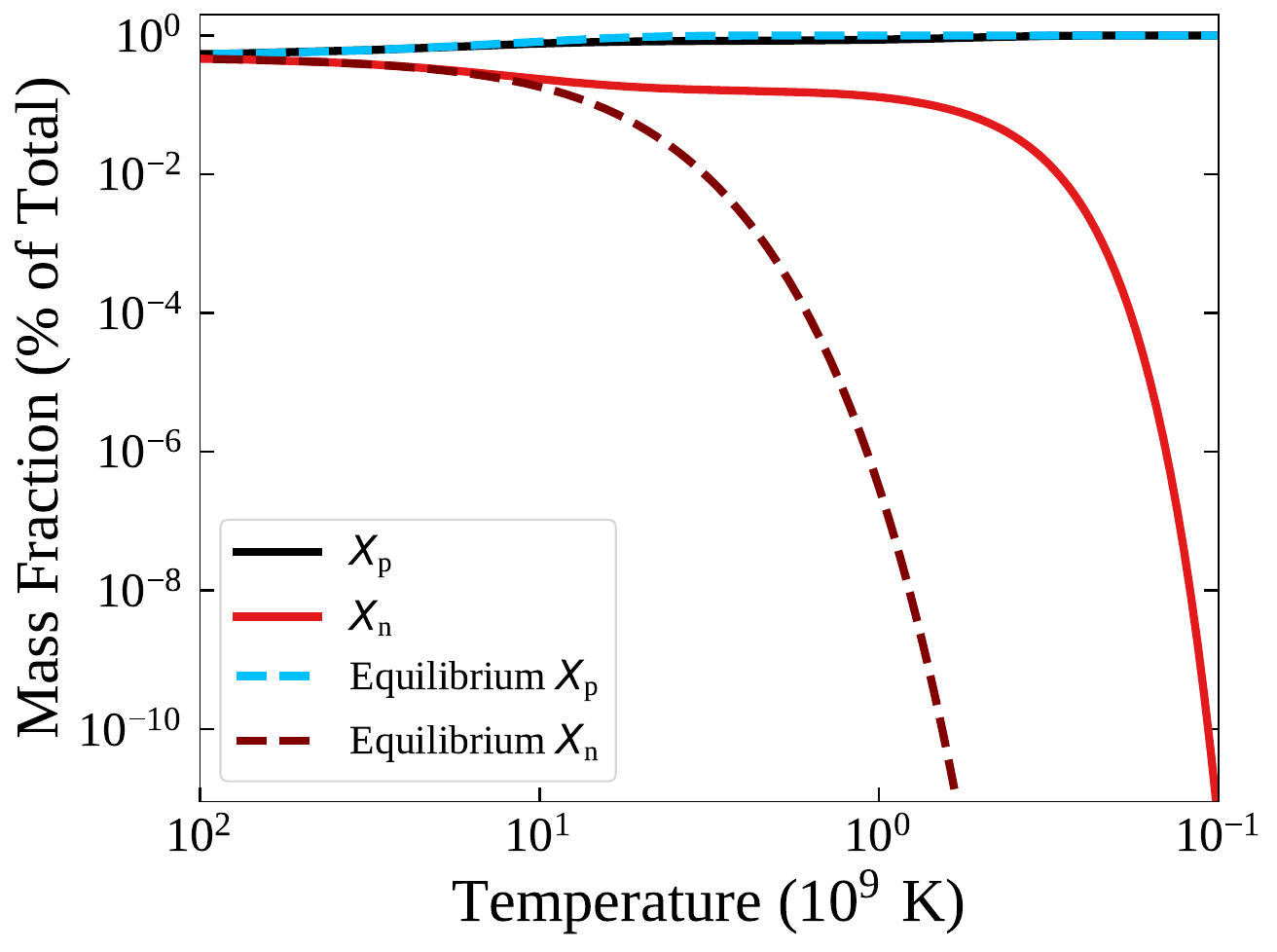}
    \caption{Proton-neutron weak interactions and their effects. \textit{Left panel:} Weak rates as a function of temperature. The black line shows the neutron-to-proton rate, while the red line shows the proton-to-neutron rate. The blue dashed line shows the inverse neutron decay time; note that the neutrons never decay slower than this limit. 
    \textit{Right panel:} mass fractions of protons and neutrons as a function of temperature. The equilibrium fractions would have followed dashed lines. Note that the (free) neutrons survive in an appreciable abundance longer than they would in equilibrium.
    }
    \label{fig:WeakRates}
\end{figure*}

The resulting weak-reaction rates for the conversion of neutrons to protons and vice versa are given in the left panel of Fig.~\ref{fig:WeakRates}. Both reaction rates decrease sharply with time (or decreasing temperature $T$), but the neutron-to-proton rate never falls below the rate corresponding to the neutron decay lifetime $\tau=880.2 ~ \rm{s}$ \cite{ParticleDataGroup:2020ssz}. 

The discussion above introduced the rates of weak interactions as a function of temperature. 
The time evolution of the mass fractions of protons and neutrons that result from these reactions are in turn given by a set of coupled differential equations:
\begin{equation}
\begin{aligned}
    \dv{X_n}{t} &= \lambda_{p \rightarrow n} X_n - \lambda_{n\rightarrow p} X_p, \\
    \dv{X_p}{t} &= \lambda_{n \rightarrow p} X_p - \lambda_{p\rightarrow n} X_n.
    \end{aligned}
    \label{eq:coupled_X}
\end{equation}
Here, the first terms on the right-hand sides indicate the creation rate of neutrons or protons, while the second terms encodes the destruction of that species.  These ODEs can be integrated to give the percentage of total mass of each species as a function of time or temperature. We will introduce a more general form of this type of ODE in the following subsection, when we talk about strong interactions.

The temporal evolution of the mass fractions of neutrons and protons is shown in the right panel of Fig.~\ref{fig:WeakRates}. It shows the neutron decoupling -- their departure from the equilibrium abundance -- at $T\simeq 10^{10}\K$.  Thereafter, the neutron abundance is much higher than it would be in equilibrium, though it eventually falls sharply as the free neutrons are incorporated in other nuclei, in reactions that we consider next.

\begin{StiffCombined}*

\textbf{Background.} The network of equations governing BBN comprises a set of coupled first-order linear ordinary differential equations. Despite the apparent simplicity of these equations, standard ordinary differential equation (ODE) solvers often struggle to handle these equations due to their ``stiff'' nature. Stiffness occurs when the solution of an equation varies significantly over the input variable. We encounter stiff equations in BBN (and, more generally, in stellar nucleosynthesis \cite{Arnett:1996ev}) because we are working around equilibrium where positive and negative terms (in e.g.\ Eq.~(\ref{eq:one_reaction_forw_back})) nearly cancel, and because reactions occur across vastly different temperature scales, with rates varying by many orders of magnitude. 
Since reducing the step size arbitrarily is highly inefficient and sometimes unfeasible, the standard \textbf{explicit methods} for solving differential equations, where a finite difference is taken to approximate the solution, are usually inadequate for stiff-equation cases. The most basic explicit method -- Euler's method -- employs a discretely sized step (denoted as $h$) to approximate the solution. For instance, Euler's method approximates the solution of the differential equation: $y'= f(y,t), \, \, y(t_0) = y_0$ is written as
\begin{equation}
y_{n+1} \approx y_n + hf(t_{n},y_{n}),
\end{equation}
where, $t_n= t_0+nh$ for $n$ steps (assuming a constant $h$). More sophisticated, multi-step explicit methods, like the Runge-Kutta method, follow this same general principle but vary in their approach. However, explicit methods are prone to instability in certain situations, as errors can propagate and lead to uncontrolled oscillations. This instability poses a challenge in solving stiff equations using explicit methods.\\

\textbf{Implicit Methods.}
Implicit methods offer an alternative approach by modifying how step sizes are evaluated. For instance, the implicit Euler method approximates the solution of the above differential equation as:
\begin{equation}
y_{n+1} \approx y_n + hf(t_{n+1},y_{n+1})
\end{equation}
Here, $f$ is evaluated at $n+1$, representing a backward differencing approach where each step looks backward in time. Although this change significantly slows down the solver, as it involves solving or approximating nonlinear systems, it drastically enhances stability. Implicit methods allow us to prevent divergence by ``looking ahead'' before taking a step, making them suitable for solving stiff equations.\\

\textbf{Worked example.} To illustrate the need for stiff integration in a simple example, consider the so-called Robertson Problem. This is a model for chemical reactions for three species $x$, $y$, and $z$, and thus not unlike the reactions for species in BBN.  An example is given by the coupled set of ODEs
\begin{equation}
\begin{cases}

\dot x &= -0.04x + 1\times10^4 y z \\
\dot y &= -0.04x - 1\times10^4 yz- 3\times10^7y^2 \\
\dot z &= 3\times10^6 y^2

\end{cases}
\end{equation}
where a dot is the derivative with respect to time.
The coefficients in this problem are many orders of magnitude different, leading to the desired property of having vastly different time-scales in the problem. We implement an explicit method (in Python, using \texttt{scipy.integrate} and \texttt{solve\_ivp}, with the Runge-Kutta RK45 method), and an implicit one (implemented via the same Python package, which relies on the Radau method).

\begin{center}
\includegraphics[width=0.4\linewidth]{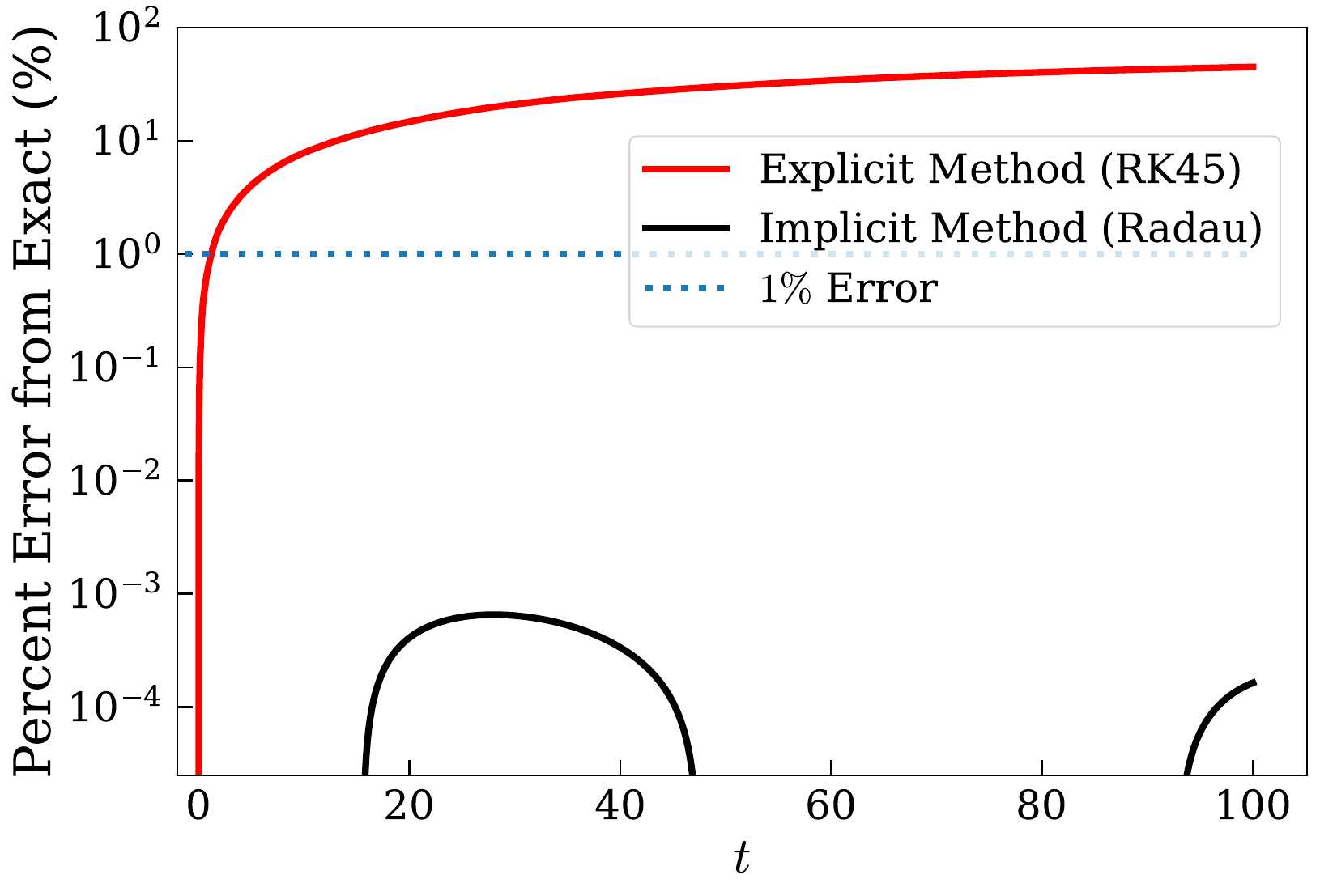}
\includegraphics[width=0.4\linewidth]{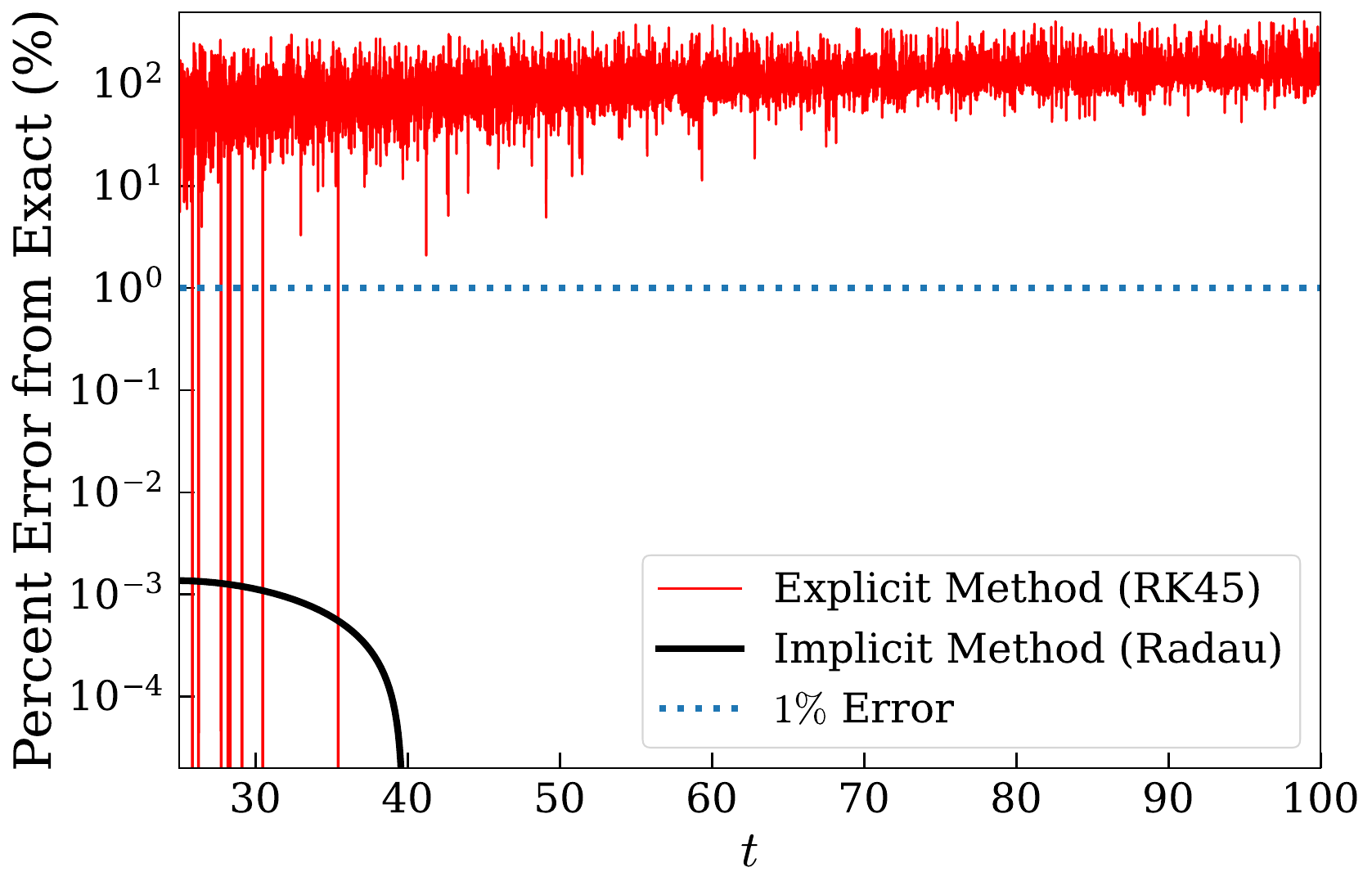} 
\end{center}

The plots above show that the explicit method (RK45) applied to this problem encounters difficulties: the solution for $x(t)$ diverges (left panel), while that for $y(t)$ oscillates (right panel); neither problem is resolved by reducing the step size arbitrarily. If, instead, we use the implicit (Radau) method, these problems disappear, and the variables quickly converge to stable values. 
\end{StiffCombined}

\subsection{Reaction Network}
\label{sec:network} 

Species begin to depart nuclear statistical equilibrium when the universe sufficiently cools; they undergo strong and electromagnetic nuclear reactions. The changes in abundances of the elements are characterized by a reaction network, represented by a set of coupled first-order nonlinear differential equations. These equations include information about the mass fraction of each species and its reaction rate as a function of temperature. The initial conditions for these differential equations are the equilibrium abundances of the species, which are the values of each $N_i$ for which the left-hand side --- and, consequently, also the right-hand side --  in Eq.~(\ref{eq:numberdensitynetwork}) is zero due to the forward and reverse rates canceling in NSE. In this section, we will build a simple yet general reaction network (which is further illustrated in Box 2), and discuss how to solve it.

\subsubsection{Nuclear Cross-Sections}
The reaction rates governing the evolution of nuclear species are essential for calculating BBN abundances, yet they are not directly measurable (as that would require doing experiments in a gas with temperature $T\sim 10^9\K$), nor can they be predicted from theory alone. Therefore, some combination of experimental and theoretical treatment is required. We now present the basics of how nuclear reactions are obtained; a more in-depth treatment is given in standard nuclear-physics textbooks such as \citet{RolfsRodney}. 

The reaction rate between two species $i$ and $j$ determines the rate at which nuclear reactions occur as a function of temperature during BBN. This rate is represented as the thermal average of the product $\sigma v$, where the reaction cross section $\sigma$ is a function of velocity (or, alternatively, energy), and $v$ is the velocity.
Given this cross-section, $\sigma_{ij}(v)$, the thermal average is given by the integral 
\begin{equation}
\langle \sigma v\rangle_{ij} = \int_0^\infty \dd v ~ \sigma_{ij}(v) v f(v),\end{equation}
where $f(v)$ is the Maxwell-Boltzmann distribution, 
\begin{equation}
    f(v) = 4\pi \left(\frac{\mu}{2\pi kT}\right)^{3/2} \exp(-\frac{\mu v^2}{2kT}),
\end{equation}
and where $\mu$ is the center of mass of the interaction. In the non-relativistic limit appropriate in this case, energy and velocity can be related via $E=\mu v^2/2$. With a change of variables from $v$ to $E$, we can alternatively write
\begin{equation}
  \!\!\langle \sigma v\rangle_{ij} = \left(\frac{8}{\pi \mu}\right)^2 \!\frac{1}{(kT)^{3/2}} \int_0^\infty \dd E\, \sigma_{ij}(E) E \exp(-\frac{E}{kT}). 
  \label{eq:sigmaE}
\end{equation}

The experimentally determined reaction cross-section $\sigma_{ij}(E)$ is a key input to the reaction rate. This cross-section is measured over a range of  energies, and interpolated in between the measured values. Using the measured cross-section values, the reaction rate as a function of temperature can be obtained from Eq.~(\ref{eq:sigmaE}). Since all the forward reactions have corresponding reverse reactions, they must also be computed, and are found generally in the same way. In practice, these integrals are often approximated as polynomial fits as a function of temperature. They can be found, for example, in the JINA ReacLib database \cite{Cyburt:2010} which we have adopted in our code.

\subsubsection{Assembling the Network}
Consider the general two-body nuclear reaction; it involves four species, labeled $i,j, k$ and $l$, where $i$ and $j$ are the reactants and $k$ and $l$ are the products. This reaction can be organizationally expressed as
\begin{equation}
i + j \leftrightharpoons k + l.    
\label{eq:simplerxn}
\end{equation}
The rate of change of the number density of one of these species, say $N_i$, can be expressed as
\begin{equation}
    \dv{N_i}{t} = N_iN_j\langle \sigma v\rangle_{ij,kl}  - N_k N_l \langle \sigma v\rangle_{kl,ij} 
    \label{eq:numberdensitynetwork}
\end{equation}
where $\langle \sigma v\rangle_{ij,kl}$ is the thermally averaged product of the reaction cross section and relative velocity in the center-of-mass system in the forward direction and, correspondingly, $\langle \sigma v\rangle_{kl,ij}$ is the same for the reverse. The quantities $\langle\sigma v\rangle $ have MKS units of length cubed per time or, in natural units (where length is equivalent to time), just length squared.

There are however some subtle corrections to Eq.~(\ref{eq:numberdensitynetwork}) when identical particles are considered. For the case when $i=j$, so that the reaction is 
\begin{equation}
i + i \leftrightharpoons k + l,  
\end{equation}
we must multiply the velocity-cross section quantity with a prefactor, as  $N_i^2/2!\, \langle \sigma v\rangle_{ii,kl}$, in order to avoid double counting. 
 The rate of change of the number density $N_i$ then becomes
\begin{equation}
    \dv{N_i}{t} = \frac{N_i^2}{2!}\langle \sigma v\rangle_{ii,kl}  - N_k N_l \langle \sigma v\rangle_{kl,ij}. 
     \label{eq:doublecount}
\end{equation}
The same logic extends to the three-body reaction. We do not consider any three-body reactions in our treatment of BBN, 
but a full treatment of BBN or stellar nucleosynthetic networks have a number of three-body reactions (such as the triple-alpha process in stellar physics). A full explanation of double and triple counting of species is offered in \citet{FCZI}. 

One more special type of reaction to consider is the one that produces one nucleus and one photon, or 
\begin{equation}
i + j \leftrightharpoons k + \gamma,
  \label{eq:photoionization}
\end{equation}
which is also known as a ``radiative capture'' reaction. In this case, the rate of change of the number density of a species, say $N_i$, is 
\begin{equation}
  \dv{N_i}{t} = N_iN_j\langle \sigma v\rangle_{ij,kl}  - N_k \langle \sigma v\rangle_{k\gamma,ij},
\end{equation}
where $\langle \sigma v\rangle_{k\gamma,ij}$ is the reverse rate. The reverse rate in this case corresponds to either $\beta$-decay, electron capture, or the photodisintegration rate, depending on the specific reaction. 

With these reactions in hand, we can construct the reaction network. By applying Eq.~(\ref{eq:massfrac}) to Eq.~(\ref{eq:numberdensitynetwork}), we can express the general four-species reaction in terms of mass fraction, $X_i$, instead of number density, $N_i$. This yields the differential equation

\begin{equation}
    \dv{X_i}{t} =  X_iX_j \rho_b N_A \langle \sigma v\rangle_{ij,kl}- X_kX_l \rho_b N_A \langle \sigma v\rangle_{kl,ij}.
\end{equation}

\begin{figure*}[t]
    \centering
    \includegraphics[width=\textwidth]{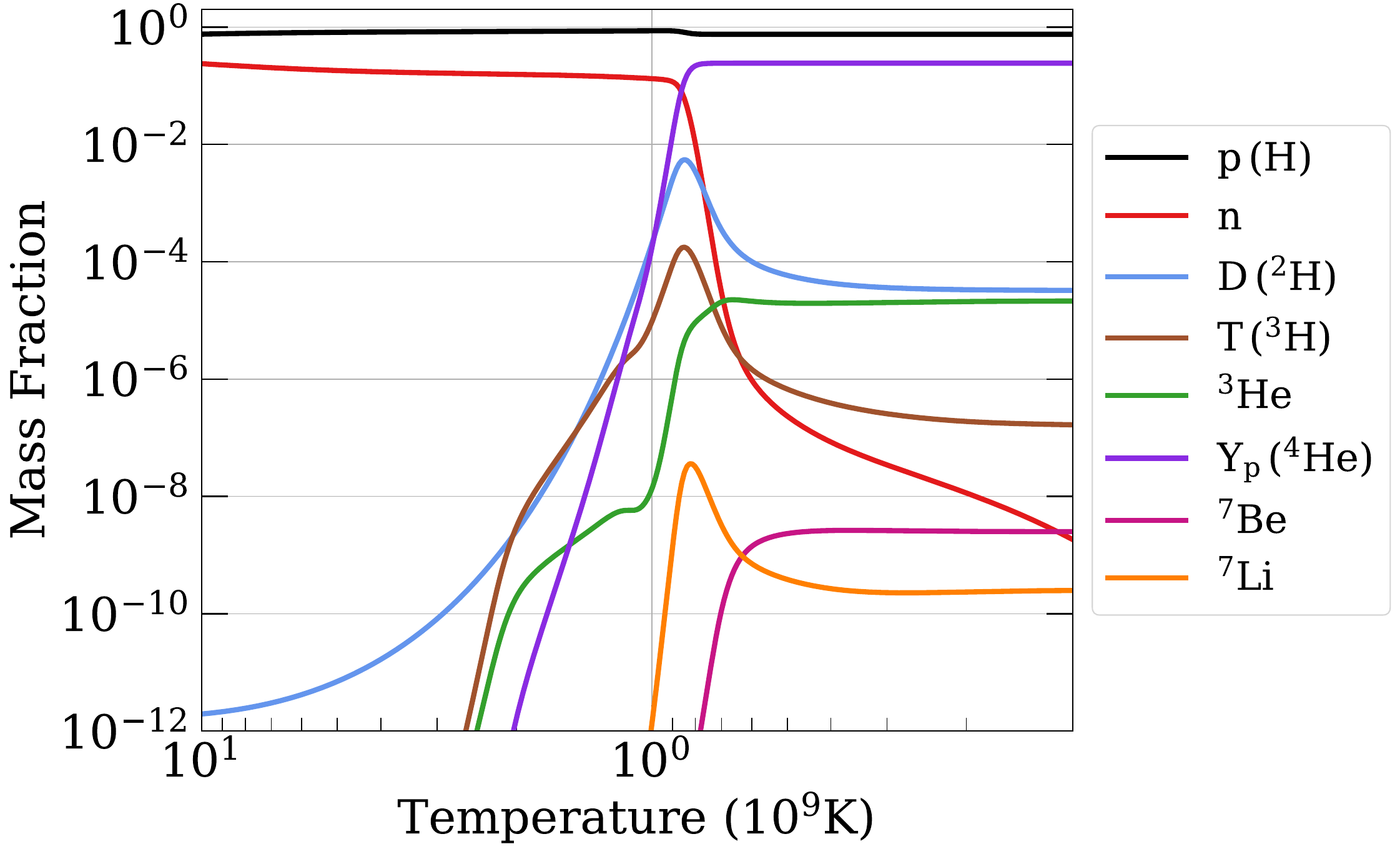}
    \caption{Abundances of the elements created during BBN, produced using \texttt{BBN-simple}. For this calculation we adopt the baryon-to-photon ratio 
$\eta = 6.12\times 10^{-10}$, the effective number of relativistic species $\Neff= 3.046$, and the mean lifetime of the neutron, $\tau_n \approx 880.2 ~ \rm{s}$. See text for other details. }
    \label{fig:Abundances}
\end{figure*}

It is customary and convenient to use the shorthand notation
\begin{equation}
[ij]_{{kl}} \equiv \rho_{b}N_{A} \langle \sigma v \rangle_{{ij}}, 
\label{eq:forwardrate}
\end{equation}
where, again, $\rho_b$ is the baryon density and $N_A$ is Avogadro's number. Conveniently, $[ij]_{{kl}}$ \footnote{This is also  expressed in some literature as the symbol $\Gamma$ (or some variation such as $\Gamma_{{ij\rightarrow kl}}$) for forwards reactions.} has units of reactions per time and thus can naturally be called the ``reaction rate'' of a given reaction.


Likewise, the reverse reaction rate, $[kl]_{{ij}}$, can be expressed in terms of the forward reaction rate $[ij]_{{kl}}$ as 

\begin{equation}
[kl]_{{ij}} = \frac{g_{i}g_{j}}{g_{k}g_{l}} \left( \frac{A_{i}A_{j}}{A_{k}A_{l}} \right)^{3/2} \exp\left( \frac{Q}{k_BT} \right) [ij]_{kl},
\label{eq:revrate}
\end{equation}
where $g_i$ is the number of spin states of each nucleus, $A_i$ is its atomic number, and $Q$ is the mass difference of the products and the reactants.

In the case of the photodissociation reaction in Eq.~(\ref{eq:photoionization}), the reverse rate $[k\gamma]_{ij}$ can be expressed as\footnote{Note that the photodissociation reactions are denoted in the literature as $\lambda_{ij}$, but we adopt the notation $[kl]_{i\gamma}$ for consistency and clarity.}

\begin{equation}
[k\gamma]_{ij} = \frac{g_{i}g_{j}}{(1+\delta_{{ij}})g_{k}} \left( \frac{A_{i}A_{j}}{A_{k}} \right)^{3/2} \rho_{b}^{-1}T_{9}^{3/2}[ij]_{k\gamma} \exp\left( \frac{Q}{k_BT}\right).
\end{equation}
The Kronecker delta function $\delta_{{ij}}$ serves in this case to avoid double counting the product $k$. Generally, the reverse reactions are far slower than the forward reactions and they contribute little to the final abundances, with the exception of $D + \gamma \rightarrow  p + n$. Nevertheless, we include them, as they are straightforward to implement. 

Rewriting Eq.~(\ref{eq:simplerxn}) in terms of these $X_i$ and using Eqs.~(\ref{eq:forwardrate}) and (\ref{eq:revrate}), we can write express the reaction equation of four species more transparently as
\begin{equation}
X_i + X_j 
\xrightleftharpoons[{[kl]_{ij}}]{\,[ij]_{kl}\,}
X_k + X_l. 
    \label{eq:one_reaction_forw_back}
\end{equation}
The differential equation for the mass fraction of a species $i$ in such a four-species reaction is 
\begin{equation}
    \dv{X_i}{t} =  -X_iX_j [ij]_{kl} +  X_kX_l [kl]_{ij}.
    \label{eq:onerxnnetwork}
\end{equation}
Similarly, in the case of photon emission, with $X_i + X_j 
\rightleftharpoons X_k + \gamma$, the differential equation for $X_i$ looks the same as Eq.~(\ref{eq:onerxnnetwork}), except with the last term's rate is proportional to $X_k$ rather than $X_kX_l$.

We are now in a position to adopt these rates and make use of the compact notation for reaction equations to create the full reaction network. The change of the mass fraction of species $i$ in a four body reaction the  can be expressed as the sum over all nuclear reactions, essentially by summing each term of Equation \ref{eq:onerxnnetwork}. The result is 
\begin{equation}
\frac{dX_i}{dt} = \sum_{j,k,l}  N_i \left(\frac{X_k^{N_k}X_l^{N_l}}{N_k!N_l!} [kl]_{ij} - \frac{X_i^{N_i}X_j^{N_j}}{N_i!N_j!} [ij]_{kl}\right)
\label{eq:coupled}
\end{equation}
Note the appearance of factorials $N_i$ which are required to avoid double counting (as discussed around Equation \ref{eq:doublecount}), while the appearance of $N_i$ in the exponents allows for the possibility of two or more identical nuclei in a reaction.  This, when written down for each species $i$, gives us a set of coupled first-order non-linear ordinary differential equations. 

Let us give a specific worked example. Consider tracking the abundance of protons. Let us only consider the reactions with protons, neutrons and deuterium (reactions 1.\ and 2.\ from Box 2)
\begin{equation}
\begin{aligned}
    p &\leftrightharpoons n \\
    p+n &\leftrightharpoons \rm{D} + \gamma,
    \label{eq:proton_reactions}
\end{aligned}
\end{equation}
and ignore, for simplicity, the higher-order reactions.
[Note that the first reaction above represents all six of the weak proton-neutron reactions from Eq.~({\ref{eq:weak_rxns}).] We then have a system of three coupled differential equations, one for each species. Applied to protons' reactions in Eq.~(\ref{eq:proton_reactions}), Eq.~(\ref{eq:coupled}) becomes
\begin{equation}
\begin{aligned}
    \dv{X_p}{t} &= - [p]_{n}  X_{n} +  [n]_{p} X_{p} \\
    & -X_{D}[np]_{D\gamma} + X_nX_p  [D\gamma]_{np},  
    \label{eq:dXp_dt}
\end{aligned}
\end{equation}
where, to link the bracket notation to commonly adopted one in weak reactions, $[p]_n \equiv \lambda_{p\rightarrow n}$ and $[n]_p\equiv \lambda_{n \rightarrow p}$.
The first line in Eq.~(\ref{eq:dXp_dt}) corresponds to the first reaction in Eq.~(\ref{eq:proton_reactions}), where the two terms in the former equation respectively account for the rate at which the protons are destroyed or created. The same is true for the respective second lines these two equations.

Similar equations can be written for neutrons and deuterium.

\renewcommand{\arraystretch}{1.20}
\begin{table*}[t]
\begin{ruledtabular}
\begin{tabular}{cccccc}
   \centering
    \textsc{Species} & \textsc{\textbf{Our Results}} & \textsc{Kawano} & \textsc{PRIMAT} & \textsc{AlterBBN} & \textsc{PArthENoPE}\\ \midrule
    H (p) & \textbf{0.7579} & 0.7530 & 0.7528 & 0.7526 & 0.7543\\
    $Y_{\rm p} (^4{\rm He}$) & \textbf{0.2420} & 0.2466 & 0.2471 & 0.2473&0.2469\\
    D/H $\times 10^5$ & \textbf{2.531} & 2.569 & 2.459 & 2.422&2.525\\
    $^3 {\rm He}$/H $\times 10^5$ & \textbf{1.019} &1.035 & 1.066 & 1.029 &1.034\\
    T/H $\times 10^8$ & \textbf{8.441} &8.171 &  7.961 & 7.654& 8.169\\
    ($^7 {\rm Be}$ + $^7{\rm Li}$)/H $\times 10^{10}$ & \textbf{5.319} & 4.509 & 5.381 & 5.308 & 4.381\\
\end{tabular}
\end{ruledtabular}
\caption{Final element abundances in our fiducial calculation (2nd column), compared to outputs by the historically influential Kawano code \cite{Kawano:1992ua}, and modern codes \texttt{Primat} \cite{Pitrou:2018cgg}, \texttt{AlterBBN} \cite{Arbey:2011nf}, and \texttt{PArthENoPE} \cite{Pisanti:2007hk}. For all calculations, we have assumed a consistent, fixed choice of the baryon-to-photon density and other relevant input parameters, as discussed in the text. }
\label{tab:abund_accuracy}
\end{table*}

\subsection{Nuclear Interactions}
\label{sec:strong}

The final piece to solve for the mass abundances in BBN is specifying the thermally averaged nuclear interaction cross sections $\langle \sigma v\rangle_{ij, kl}$ introduced in Section \ref{sec:network}, which are required for the reaction network. As mentioned around Eq.~(\ref{eq:sigmaE}), the thermally averaged cross-sections are obtained from nuclear cross-sections which are experimentally measured in accelerators at a range of energies, then tabulated, and finally connected to a theoretical model that allows thermal averaging. Much of the effort to measure these cross-sections and determine nuclear rates has been motivated by studies of stellar nucleosynthesis, as well as fusion energy and nuclear weapons research.

Here we adopt the rates from the ReacLib database \cite{Cyburt:2010}, and  ignore any uncertainties in them. This choice provides a straightforward one-stop-shop for the rates. Our emphasis on simplicity rather than precision justifies the ignorance of any updated best-fit values in these rates, or the uncertainties in them (see e.g.~\cite{Krauss:1989np} for early work where the rate uncertainties were incorporated). 
For a comprehensive historical overview of rate computation and the nuclear physics involved, see \citet{AnguloArnould99}.

The primary reactions between nuclei are shown in Box 2, which also includes diagrams visually illustrating these reactions and showing their rates as a function of temperature.


\section{Summary and Results}
\label{sec:results}

To summarize the procedure described in the paper thus far, computation of BBN element abundances requires the following actions: 
\begin{itemize}
    \item Derive the time-temperature relationship (Eq.~(\ref{eq:Tt-relation})) in the early universe by making use of the Friedmann equations (Eqs.~(\ref{eq:FI}) and (\ref{eq:FII})), as well as the continuity equation (Eq.~(\ref{eq:continuity})). This calculation requires the energy densities and pressures of relevant species (photons, neutrinos, electrons, and positrons).
    \item Derive the relationship between the photon and neutrino temperatures, $T_\gamma$ and $T_\nu$, that is governed by the decoupling of neutrinos, electrons, and positrons from photons around $T \gg m_e$ (Eq.~(\ref{eq:PhotonNeutrinoRelationship})).
    \item Set up the nuclear statistical equilibrium following Eq.~(\ref{eq:nse}) at some temperature $T\gg 1\MeV$. In addition to the temperature (and the species' masses and binding energies), the required input is the baryon density or, equivalently, the baryon-to-photon number $\eta_b$ .
    \item Set up the weak rates between the photons and neutrons resulting in the integrals in Eq.~(\ref{eq:weak_rates}) and Eq.~(\ref{eq:rev_weak}). Due to the complexity of the integral, it is necessary to implement the numerical methods described in Box 1. 
    \item Evaluate the forwards and reverse rates $[kl]_{{ij}}$ and $[ij]_{kl}$; see Eqs.~(\ref{eq:forwardrate}) and (\ref{eq:revrate}).
    \item Evolve the mass fractions of species, $X_i$, in time. The general equation that governs this is Eq.~(\ref{eq:coupled}), with details and possible simplifications as discussed in Sec.~\ref{sec:network}. The required input are the forwards and reverse rates, as well as the photon and neutrino temperatures and the time-temperature equation. Because the coupled ODEs are ``stiff'' as they involve vastly different timescales, specialized techniques described in Box 3 are recommended.
\end{itemize}

We do this for some of the lightest nuclei (H, $^4$He, D, T, $^3$He, $^7$Li, $^7$Be) and consider the 12 principal reactions listed in Box 2. For our fiducial calculation, we adopt the baryon-to-photon ratio 
$\eta = 6.12\times 10^{-10}$ \cite{Planck:2018vyg}, which corresponds to physical baryon density $\Omega_bh^2=0.0224$, which is consistent with the measurements from the temperature and polarization anisotropies in the CMB when compared to the standard cosmological model \cite{Planck:2018vyg} and the deuterium abundance compared to BBN theory (e.g.~\cite{Mossa:2020gjc}). We also assume the effective number of relativistic species $\Neff= 3.046$ recall from footnote\footref{foot:Neff} that $\Neff$ affects the expansion rate during BBN, and hence the elemental abundances) and the mean lifetime of the neutron $\tau_n \approx 880.2\, \rm{s}$ \cite{ParticleDataGroup:2020ssz}.

The results of this calculation, with steps and assumptions as described above, are shown in Fig.~\ref{fig:Abundances}. It shows the familiar results of BBN in the hot Big Bang cosmological framework: dominant fractions of hydrogen and helium, smaller abundances of deuterium, tritium, helium-3, trace amounts of lithium-7 and beryllium, and a rapidly vanishing abundance of free neutrons. [We follow a common practice to show only the sum of the abundances of lithium-7 and beryllium-7, because $^7$Be quickly (with a lifetime of about 53 days) decays into $^7$Li, so that only the sum is observable today\footnote{The same is actually true of helium-3 and tritium, as T decays into $^3$He in about 12 years; we however choose to show the abundances of these two species separately.}.] 

Our calculation also reflects familiar dependencies of the elements' abundance on the physical conditions during BBN (which we do not show separately, but find it easy to verify with our code). A faster exit from weak equilibrium  would imply a higher neutron-to-proton ratio (as the neutrons have less time to beta-decay to protons; see the right panel of Fig.~\ref{fig:WeakRates}), and hence a higher final $^4$He abundance. A faster expansion rate $H$ during BBN (for a fixed baryon density) that is enabled, for example, by higher value of $\Neff$ or a lower $\eta$ (both indicating more radiation density), would lead to less time to burn deuterium, and hence a larger observed abundance of deuterium. 

We next discuss the accuracy of our (admittedly very simple!) calculation, and compare it to professional BBN codes. This is shown in Table \ref{tab:abund_accuracy}. To do so, we ran the leading BBN codes \texttt{Primat} \cite{Pitrou:2018cgg}, \texttt{AlterBBN} \cite{Arbey:2011nf}, and \texttt{PArthENoPE} \cite{Pisanti:2007hk}; we have assumed the same values of $\eta_b$, $\Neff$, and $\tau_n$ as in Fig.~\ref{fig:Abundances} that were listed just above. The Table shows that,  despite our explicitly pedagogical approach not aimed at high precision, the results are in a reasonably good agreement with the other codes: our helium-4 abundance agrees to about a percent, the deuterium and helium-3 abundances are accurate to a few percent, and the tritium abundance is accurate to within 10\%. These accuracies are respectable given that we have ignored quantum corrections and adopted approximations at several stages of the analysis. It is also noteworthy that there are non-negligible mutual differences in the abundances computed by the other three codes as well. Given that the three modern professional-grade codes (\texttt{Primat}, \texttt{AlterBBN}, and \texttt{PArthENoPE}) use newer rates and more robust statistical analyses, the accuracy of our code seems respectable, and is sufficient for many basic applications in cosmology. 


\section{Conclusions}
\label{sec:concl} 

We have presented a pedagogical, from-scratch numerical calculation of the abundances of lightest elements created during the big bang nucleosynthesis (BBN). We outlined a step-by-step procedure to establish all required quantities in order to carry out the BBN calculation. The code that we developed, \texttt{BBN-simple}, is made available in a transparent and easy-to-use graphical user interface.

The principal ingredients to BBN include setting up the thermodynamical relations between temperature, time, and energy densities of different species (discussed in Sec.~\ref{sec:thermo}), and establishing the temperature dependence of the various relevant weak and strong nuclear reactions (discussed in Sec.~\ref{sec:reactions}).  The results, presented in Sec.~\ref{sec:results}, culminate in the familiar plot of 
evolution of the abundances in cosmic time (or decreasing temperature), shown in Fig.~\ref{fig:Abundances} for the input parameters of the standard cosmological model. 

We also discussed in some depth specific challenges that a student/researcher would encounter in setting up the BBN calculation. In Box 1, we discuss a simple and effective way to implement weak-interaction rates into the calculation. In Box 2, we lay out a basic BBN reaction network, with references to where the rates can be found. Perhaps the most non-trivial part of the calculation is the ``stiff'' nature of ordinary differential equations in the reaction network; computational techniques to overcome this are discussed in Box 3.  The accuracy of our calculation is also studied, and we present basic comparisons in Table \ref{tab:abund_accuracy}. We find that the accuracy is reasonably good --- percent-level for helium, deuterium, and helium-3 for example. The accuracy of the final abundances could be further improved with the use of  modern, experimentally measured nuclear-reaction rates, but we do not do so in this paper. 

We hope that this presentation will be useful to students who would like to write their own BBN code from scratch, or else to beginning researchers in the field who would like to get hands-on experience and who wish to better understand the high-precision codes that are on the market. We have provided numerical code and basic explanations at \url{http://www-personal.umich.edu/~aidanmw/}.

\acknowledgements

We thank Ken Nollett for many useful comments on an earlier version of this manuscript. 

\newpage 

\appendix






\bibliography{references}

@BOOK{1942psd..book.....C,
       author = {{Chandrasekhar}, Subrahmanyan},
        title = "{Principles of stellar dynamics}",
         year = 1942,
       adsurl = {https://ui.adsabs.harvard.edu/abs/1942psd..book.....C}
}

@article{Alpher:1948ve,
    author = "Alpher, R. A. and Bethe, H. and Gamow, G.",
    title = "{The origin of chemical elements}",
    doi = "10.1103/PhysRev.73.803",
    journal = "Phys. Rev.",
    volume = "73",
    pages = "803--804",
    year = "1948"
}

@article{Gamow:1948pob,
    author = "Gamow, G.",
    title = "{The Evolution of the Universe}",
    doi = "10.1038/162680a0",
    journal = "Nature",
    volume = "162",
    number = "4122",
    pages = "680--682",
    year = "1948"
}

@article{Alpher:1948srz,
    author = "Alpher, Ralph A. and Herman, Robert",
    title = "{Evolution of the Universe}",
    doi = "10.1038/162774b0",
    journal = "Nature",
    volume = "162",
    number = "4124",
    pages = "774--775",
    year = "1948"
}

@article{Alpher:1948yqy,
    author = "Alpher, R. A.",
    title = "{A Neutron-Capture Theory of the Formation and Relative Abundance of the Elements}",
    doi = "10.1103/PhysRev.74.1577",
    journal = "Phys. Rev.",
    volume = "74",
    number = "11",
    pages = "1577--1589",
    year = "1948"
}

@article{FCZI,
       author = {Fowler, William A. and Caughlan, Georgeanne R. and Zimmerman, Barbara A.},
        title = "{Thermonuclear Reaction Rates}",
      journal = {\araa},
         year = 1967,
        month = jan,
       volume = {5},
        pages = {525},
          doi = {10.1146/annurev.aa.05.090167.002521},
       adsurl = {https://ui.adsabs.harvard.edu/abs/1967ARA&A...5..525F}
}

@article{FCZII,
       author = {{Fowler}, William A. and {Caughlan}, Georgeanne R. and {Zimmerman}, Barbara A.},
        title = "{Thermonuclear Reaction Rates, II}",
      journal = {\araa},
         year = 1975,
        month = jan,
       volume = {13},
        pages = {69},
          doi = {10.1146/annurev.aa.13.090175.000441},
       adsurl = {https://ui.adsabs.harvard.edu/abs/1975ARA&A..13...69F}
}

@article{FCZIII,
       author = {{Harris}, M.~J. and {Fowler}, W.~A. and {Caughlan}, G.~R. and {Zimmerman}, B.~A.},
        title = "{Thermonuclear reaction rates, III.}",
      journal = {\araa},
         year = 1983,
        month = jan,
       volume = {21},
        pages = {165-176},
          doi = {10.1146/annurev.aa.21.090183.001121},
       adsurl = {https://ui.adsabs.harvard.edu/abs/1983ARA&A..21..165H}
}

@article{FCZIV,
       author = {{Caughlan}, G.~R. and {Fowler}, W.~A. and {Harris}, M.~J. and {Zimmerman}, B.~A.},
        title = "{Tables of Thermonuclear Reaction Rates for Low-Mass Nuclei ($1 \leq Z \leq 14$)}",
      journal = {Atomic Data and Nuclear Data Tables},
         year = 1985,
        month = jan,
       volume = {32},
        pages = {197},
          doi = {10.1016/0092-640X(85)90006-3},
       adsurl = {https://ui.adsabs.harvard.edu/abs/1985ADNDT..32..197C}
}

@article{FCZV,
       author = {{Caughlan}, Georgeann R. and {Fowler}, William A.},
        title = "{Thermonuclear Reaction Rates V}",
      journal = {Atomic Data and Nuclear Data Tables},
         year = 1988,
        month = jan,
       volume = {40},
        pages = {283},
          doi = {10.1016/0092-640X(88)90009-5},
       adsurl = {https://ui.adsabs.harvard.edu/abs/1988ADNDT..40..283C}
}

@article{Wagoner67,
       author = {{Wagoner}, Robert V. and {Fowler}, William A. and {Hoyle}, F.},
        title = "{On the Synthesis of Elements at Very High Temperatures}",
      journal = {\apj},
         year = 1967,
        month = apr,
       volume = {148},
        pages = {3},
          doi = {10.1086/149126},
       adsurl = {https://ui.adsabs.harvard.edu/abs/1967ApJ...148....3W}
}

@article{Hayashi:1950lqo,
    author = "Hayashi, C.",
    title = "{Proton-Neutron Concentration Ratio in the Expanding Universe at the Stages preceding the Formation of the Elements}",
    doi = "10.1143/ptp/5.2.224",
    journal = "Prog. Theor. Phys.",
    volume = "5",
    number = "2",
    pages = "224--235",
    year = "1950"
}

@article{Peebles:1966zz,
    author = "Peebles, P. J. E.",
    title = "{Primordial Helium Abundance and the Primordial Fireball. 2}",
    doi = "10.1086/148918",
    journal = "Astrophys. J.",
    volume = "146",
    pages = "542--552",
    year = "1966"
}

@article{Alpher:1953zz,
    author = "Alpher, Ralph A. and Follin, James W. and Herman, Robert C.",
    title = "{Physical Conditions in the Initial Stages of the Expanding Universe}",
    doi = "10.1103/PhysRev.92.1347",
    journal = "Phys. Rev.",
    volume = "92",
    pages = "1347--1361",
    year = "1953"
}

@ARTICLE{FowlerHoyle1964,
       author = {{Fowler}, William A. and {Hoyle}, F.},
        title = "{Neutrino Processes and Pair Formation in Massive Stars and Supernovae.}",
      journal = {\apjs},
         year = 1964,
        month = dec,
       volume = {9},
        pages = {201},
          doi = {10.1086/190103},
       adsurl = {https://ui.adsabs.harvard.edu/abs/1964ApJS....9..201F}
}

@ARTICLE{1983MNRAS.205..683S,
       author = {{Scherrer}, R.~J.},
        title = "{Primordial element production in universes with large lepton-baryon ratio}",
      journal = {\mnras},
         year = 1983,
        month = nov,
       volume = {205},
        pages = {683-690},
          doi = {10.1093/mnras/205.3.683},
       adsurl = {https://ui.adsabs.harvard.edu/abs/1983MNRAS.205..683S}
}

@article{AnguloArnould99,
title = {A compilation of charged-particle induced thermonuclear reaction rates},
journal = {Nuclear Physics A},
volume = {656},
number = {1},
pages = {3-183},
year = {1999},
issn = {0375-9474},
doi = {https://doi.org/10.1016/S0375-9474(99)00030-5},
url = {https://www.sciencedirect.com/science/article/pii/S0375947499000305},
author = {C. Angulo and M. Arnould and M. Rayet and P. Descouvemont and D. Baye and C. Leclercq-Willain and A. Coc and S. Barhoumi and P. Aguer and C. Rolfs and R. Kunz and J.W. Hammer and A. Mayer and T. Paradellis and S. Kossionides and C. Chronidou and K. Spyrou and S. Degl'Innocenti and G. Fiorentini and B. Ricci and S. Zavatarelli and C. Providencia and H. Wolters and J. Soares and C. Grama and J. Rahighi and A. Shotter and M. {Lamehi Rachti}}
}

@article{Krauss:1989np,
    author = "Krauss, Lawrence M. and Romanelli, Paul",
    title = "{Big Bang Nucleosynthesis: Predictions And Uncertainties}",
    reportNumber = "YCTP-P1-89A",
    doi = "10.1086/168962",
    journal = "Astrophys. J.",
    volume = "358",
    pages = "47--59",
    year = "1990"
}

@book{Kolb:1990vq,
    author = "Kolb, Edward W. and Turner, Michael S.",
    title = "{The Early Universe}",
    reportNumber = "FERMILAB-BOOK-1990-01",
    doi = "10.1201/9780429492860",
    isbn = "978-0-201-62674-2",
    volume = "69",
    year = "1990"
}

@book{Huterer:2023mmv,
    author = "Huterer, Dragan",
    title = "{A Course in Cosmology}",
    doi = "10.1017/9781009070232",
    isbn = "978-1-00-907023-2",
    publisher = "Cambridge University Press",
    month = "3",
    year = "2023"
}

@book{Ryden, 
       place={New York, NY}, 
       title={Introduction to Cosmology}, 
       publisher={Cambridge University Press}, 
       author={Ryden, Barbara Sue}, 
       year={2017}
}

@book{Mukhanov:2005sc,
    author = "Mukhanov, V.",
    title = "{Physical Foundations of Cosmology}",
    doi = "10.1017/CBO9780511790553",
    isbn = "978-0-521-56398-7",
    publisher = "Cambridge University Press",
    address = "Oxford",
    year = "2005"
}

@book{Arnett:1996ev,
    author = "Arnett, David",
    title = "{Supernovae and Nucleosynthesis: An Investigation of the History of Matter, from the Big Bang to the Present}",
    isbn = "978-0-691-01147-9",
    publisher = "Princeton University Press",
    month = "3",
    year = "1996"
}

@BOOK{DodelsonSchmidt,
       author = {{Dodelson}, Scott and {Schmidt}, Fabian},
        title = "{Modern Cosmology}",
         year = 2020,
          doi = {10.1016/C2017-0-01943-2},
       adsurl = {https://ui.adsabs.harvard.edu/abs/2020moco.book.....D}
}

@BOOK{RolfsRodney, 
    author = {{Rolfs}, Claus and {Rodney}, William},
    title = "{Cauldrons in the Cosmos}",
    year = 1988
}

@article{Mukhanov:2003xs,
    author = "Mukhanov, Viatcheslav F.",
    editor = "Gunzig, E. and Mukhanov, Viatcheslav F. and Verdaguer, E.",
    title = "{Nucleosynthesis without a computer}",
    eprint = "astro-ph/0303073",
    archivePrefix = "arXiv",
    doi = "10.1023/B:IJTP.0000048169.69609.77",
    journal = "Int. J. Theor. Phys.",
    volume = "43",
    pages = "669--693",
    year = "2004"
}

@article{Esmailzadeh:1990hf,
    author = "Esmailzadeh, Rahim and Starkman, Glenn D. and Dimopoulos, Savas",
    title = "{Primordial nucleosynthesis without a computer}",
    reportNumber = "CFPA-TH-90-023, IASSNS-AST-90-24",
    journal = "Astrophys. J.",
    volume = "378",
    pages = "504--518",
    year = "1991"
}

@article{Turner:2021roa,
    author = "Turner, Michael S. and KICP/UChicago and Foundation, The Kavli",
    title = "{Understanding BBN: the physics and its history}",
    eprint = "2111.14254",
    journal = "",
    archivePrefix = "arXiv",
    primaryClass = "astro-ph.CO",
    month = "11",
    year = "2021"
}

@article{Copi:1994ev,
    author = "Copi, Craig J. and Schramm, David N. and Turner, Michael S.",
    title = "{Big bang nucleosynthesis and the baryon density of the universe}",
    eprint = "astro-ph/9407006",
    archivePrefix = "arXiv",
    reportNumber = "FERMILAB-PUB-94-174-A",
    doi = "10.1126/science.7809624",
    journal = "Science",
    volume = "267",
    pages = "192--199",
    year = "1995"
}

@article{Walker:1991ap,
    author = "Walker, Terry P. and Steigman, Gary and Schramm, David N. and Olive, Keith A. and Kang, Ho-Shik",
    title = "{Primordial nucleosynthesis redux}",
    reportNumber = "OSU-TA-10-90, UMN-TH-826-90, FERMILAB-PUB-91-036-A",
    doi = "10.1086/170255",
    journal = "Astrophys. J.",
    volume = "376",
    pages = "51--69",
    year = "1991"
}

@article{Olive:1999ij,
    author = "Olive, Keith A. and Steigman, Gary and Walker, Terry P.",
    title = "{Primordial nucleosynthesis: Theory and observations}",
    eprint = "astro-ph/9905320",
    archivePrefix = "arXiv",
    reportNumber = "UMN-TH-1802-99, TPI-MINN-99-29",
    doi = "10.1016/S0370-1573(00)00031-4",
    journal = "Phys. Rept.",
    volume = "333",
    pages = "389--407",
    year = "2000"
}

@article{Cyburt:2015mya,
    author = "Cyburt, Richard H. and Fields, Brian D. and Olive, Keith A. and Yeh, Tsung-Han",
    title = "{Big Bang Nucleosynthesis: 2015}",
    eprint = "1505.01076",
    archivePrefix = "arXiv",
    primaryClass = "astro-ph.CO",
    reportNumber = "UMN-TH-3432-15, FTPI-MINN-15-19",
    doi = "10.1103/RevModPhys.88.015004",
    journal = "Rev. Mod. Phys.",
    volume = "88",
    pages = "015004",
    year = "2016"
}

@article{Kawano:1992ua,
    author = "Kawano, Lawrence",
    title = "{Let's go: Early universe. 2. Primordial nucleosynthesis: The Computer way}",
    reportNumber = "FERMILAB-PUB-92-004-A",
    journal = "",
    month = "1",
    year = "1992"
}

@ARTICLE{SmithKawanoMalaney93,
       author = {{Smith}, Michael S. and {Kawano}, Lawrence H. and {Malaney}, Robert A.},
        title = "{Experimental, Computational, and Observational Analysis of Primordial Nucleosynthesis}",
      journal = {\apjs},
         year = 1993,
        month = apr,
       volume = {85},
        pages = {219},
          doi = {10.1086/191763},
       adsurl = {https://ui.adsabs.harvard.edu/abs/1993ApJS...85..219S}
}

@article{Pisanti:2007hk,
    author = "Pisanti, O. and Cirillo, A. and Esposito, S. and Iocco, F. and Mangano, G. and Miele, G. and Serpico, P. D.",
    title = "{PArthENoPE: Public Algorithm Evaluating the Nucleosynthesis of Primordial Elements}",
    eprint = "0705.0290",
    archivePrefix = "arXiv",
    primaryClass = "astro-ph",
    reportNumber = "DSF-13-07, FERMILAB-PUB-07-079-A, SLAC-PUB-12488",
    doi = "10.1016/j.cpc.2008.02.015",
    journal = "Comput. Phys. Commun.",
    volume = "178",
    pages = "956--971",
    year = "2008"
}

@article{Arbey:2011nf,
    author = "Arbey, Alexandre",
    title = "{AlterBBN: A program for calculating the BBN abundances of the elements in alternative cosmologies}",
    eprint = "1106.1363",
    archivePrefix = "arXiv",
    primaryClass = "astro-ph.CO",
    reportNumber = "CERN-PH-TH-2011-135, LYCEN-2011-06",
    doi = "10.1016/j.cpc.2012.03.018",
    journal = "Comput. Phys. Commun.",
    volume = "183",
    pages = "1822--1831",
    year = "2012"
}

@article{Arbey:2018zfh,
    author = "Arbey, A. and Auffinger, J. and Hickerson, K. P. and Jenssen, E. S.",
    title = "{AlterBBN v2: A public code for calculating Big-Bang nucleosynthesis constraints in alternative cosmologies}",
    eprint = "1806.11095",
    archivePrefix = "arXiv",
    primaryClass = "astro-ph.CO",
    reportNumber = "CERN-TH-2018-146",
    doi = "10.1016/j.cpc.2019.106982",
    journal = "Comput. Phys. Commun.",
    volume = "248",
    pages = "106982",
    year = "2020"
}

@article{Pitrou:2018cgg,
    author = "Pitrou, Cyril and Coc, Alain and Uzan, Jean-Philippe and Vangioni, Elisabeth",
    title = "{Precision big bang nucleosynthesis with improved Helium-4 predictions}",
    eprint = "1801.08023",
    archivePrefix = "arXiv",
    primaryClass = "astro-ph.CO",
    doi = "10.1016/j.physrep.2018.04.005",
    journal = "Phys. Rept.",
    volume = "754",
    pages = "1--66",
    year = "2018"
}

@article{DiValentino:2021izs,
    author = "Di Valentino, Eleonora and Mena, Olga and Pan, Supriya and Visinelli, Luca and Yang, Weiqiang and Melchiorri, Alessandro and Mota, David F. and Riess, Adam G. and Silk, Joseph",
    title = "{In the realm of the Hubble tension\textemdash{}a review of solutions}",
    eprint = "2103.01183",
    archivePrefix = "arXiv",
    primaryClass = "astro-ph.CO",
    reportNumber = "IPPP/20/108",
    doi = "10.1088/1361-6382/ac086d",
    journal = "Class. Quant. Grav.",
    volume = "38",
    number = "15",
    pages = "153001",
    year = "2021"
}

@article{DES:2017txv,
    author = "Abbott, T. M. C. and others",
    collaboration = "DES",
    title = "{Dark Energy Survey Year 1 Results: A Precise H0 Estimate from DES Y1, BAO, and D/H Data}",
    eprint = "1711.00403",
    archivePrefix = "arXiv",
    primaryClass = "astro-ph.CO",
    reportNumber = "FERMILAB-PUB-17-482-AE",
    doi = "10.1093/mnras/sty1939",
    journal = "Mon. Not. Roy. Astron. Soc.",
    volume = "480",
    number = "3",
    pages = "3879--3888",
    year = "2018"
}

@article{Planck:2018vyg,
    author = "Aghanim, N. and others",
    collaboration = "Planck",
    title = "{Planck 2018 results. VI. Cosmological parameters}",
    eprint = "1807.06209",
    archivePrefix = "arXiv",
    primaryClass = "astro-ph.CO",
    doi = "10.1051/0004-6361/201833910",
    journal = "Astron. Astrophys.",
    volume = "641",
    pages = "A6",
    year = "2020",
    note = "[Erratum: Astron.Astrophys. 652, C4 (2021)]"
}

@article{DESI:2024mwx,
    author = "Adame, A. G. and others",
    collaboration = "DESI",
    title = "{DESI 2024 VI: Cosmological Constraints from the Measurements of Baryon Acoustic Oscillations}",
    journal = "",
    eprint = "2404.03002",
    archivePrefix = "arXiv",
    primaryClass = "astro-ph.CO",
    reportNumber = "FERMILAB-PUB-24-0154-PPD",
    month = "4",
    year = "2024"
}

@article{ParticleDataGroup:2020ssz,
    author = "Zyla, P. A. and others",
    collaboration = "Particle Data Group",
    title = "{Review of Particle Physics}",
    doi = "10.1093/ptep/ptaa104",
    journal = "PTEP",
    volume = "2020",
    number = "8",
    pages = "083C01",
    year = "2020"
}

@article{Pettini:2001yu,
    author = "Pettini, Max and Bowen, David V.",
    title = "{A new measurement of the primordial abundance of deuterium: toward convergence with the baryon density from the cmb?}",
    eprint = "astro-ph/0104474",
    archivePrefix = "arXiv",
    doi = "10.1086/322510",
    journal = "Astrophys. J.",
    volume = "560",
    pages = "41--48",
    year = "2001"
}

@article{Fumagalli:2011iw,
    author = "Fumagalli, Michele and O'Meara, John M. and Prochaska, J. Xavier",
    title = "{Detection of Pristine Gas Two Billion Years after the Big Bang}",
    eprint = "1111.2334",
    archivePrefix = "arXiv",
    primaryClass = "astro-ph.CO",
    doi = "10.1126/science.1213581",
    journal = "Science",
    volume = "334",
    pages = "1245",
    year = "2011"
}

@article{Noterdaeme:2012pa,
    author = "Noterdaeme, P. and Lopez, S. and Dumont, V. and Ledoux, C. and Molaro, P. and Petitjean, P.",
    title = "{Deuterium at high-redshift: Primordial abundance in the zabs = 2.621 damped Ly-alpha system towards CTQ247}",
    eprint = "1205.3777",
    archivePrefix = "arXiv",
    primaryClass = "astro-ph.CO",
    doi = "10.1051/0004-6361/201219453",
    journal = "Astron. Astrophys.",
    volume = "542",
    pages = "L33",
    year = "2012"
}

@article{Izotov:2014fga,
    author = "Izotov, Y. I. and Thuan, T. X. and Guseva, N. G.",
    title = "{A new determination of the primordial He abundance using the He i $\lambda$10830 \r{A} emission line: cosmological implications}",
    eprint = "1408.6953",
    archivePrefix = "arXiv",
    primaryClass = "astro-ph.CO",
    doi = "10.1093/mnras/stu1771",
    journal = "Mon. Not. Roy. Astron. Soc.",
    volume = "445",
    number = "1",
    pages = "778--793",
    year = "2014"
}

@article{Aver:2015iza,
    author = "Aver, Erik and Olive, Keith A. and Skillman, Evan D.",
    title = "{The effects of He I \ensuremath{\lambda}10830 on helium abundance determinations}",
    eprint = "1503.08146",
    archivePrefix = "arXiv",
    primaryClass = "astro-ph.CO",
    doi = "10.1088/1475-7516/2015/07/011",
    journal = "JCAP",
    volume = "07",
    pages = "011",
    year = "2015"
}

@article{Cooke:2017cwo,
    author = "Cooke, Ryan J. and Pettini, Max and Steidel, Charles C.",
    title = "{One Percent Determination of the Primordial Deuterium Abundance}",
    eprint = "1710.11129",
    archivePrefix = "arXiv",
    primaryClass = "astro-ph.CO",
    doi = "10.3847/1538-4357/aaab53",
    journal = "Astrophys. J.",
    volume = "855",
    number = "2",
    pages = "102",
    year = "2018"
}

@article{Mossa:2020gjc,
author = "Mossa, V. and others",
title = "{The baryon density of the Universe from an improved rate of deuterium burning}",
doi = "10.1038/s41586-020-2878-4",
journal = "Nature",
volume = "587",
number = "7833",
pages = "210--213",
year = "2020"
}

@article{Cyburt:2010,
doi = {10.1088/0067-0049/189/1/240},
url = {https://dx.doi.org/10.1088/0067-0049/189/1/240},
year = {2010},
month = {jun},
publisher = {The American Astronomical Society},
volume = {189},
number = {1},
pages = {240},
author = {Richard H. Cyburt and A. Matthew Amthor and Ryan Ferguson and Zach Meisel and Karl Smith and Scott Warren and Alexander Heger and R. D. Hoffman and Thomas Rauscher and Alexander Sakharuk and Hendrik Schatz and F. K. Thielemann and Michael Wiescher},
title = {THE JINA REACLIB DATABASE: ITS RECENT UPDATES AND IMPACT ON TYPE-I X-RAY BURSTS},
journal = {The Astrophysical Journal Supplement Series}
}

@article{Nollett:2011aa,
    author = "Nollett, Kenneth M. and Holder, Gilbert P.",
    title = "{An analysis of constraints on relativistic species from primordial nucleosynthesis and the cosmic microwave background}",
    eprint = "1112.2683",
    journal = "",
    archivePrefix = "arXiv",
    primaryClass = "astro-ph.CO",
    month = "12",
    year = "2011"
}

@article{Schoneberg:2024ifp,
    author = {Sch\"oneberg, Nils},
    title = "{The 2024 BBN baryon abundance update}",
    eprint = "2401.15054",
    archivePrefix = "arXiv",
    primaryClass = "astro-ph.CO",
    doi = "10.1088/1475-7516/2024/06/006",
    journal = "JCAP",
    volume = "06",
    pages = "006",
    year = "2024"
}

@article{Adelberger:2010qa,
    author = "Adelberger, E. G. and others",
    title = "{Solar fusion cross sections II: the pp chain and CNO cycles}",
    eprint = "1004.2318",
    archivePrefix = "arXiv",
    primaryClass = "nucl-ex",
    reportNumber = "INT-PUB-10-016, UCB-NPAT-10-001",
    doi = "10.1103/RevModPhys.83.195",
    journal = "Rev. Mod. Phys.",
    volume = "83",
    pages = "195",
    year = "2011"
}

@article{Iocco:2008va,
    author = "Iocco, Fabio and Mangano, Gianpiero and Miele, Gennaro and Pisanti, Ofelia and Serpico, Pasquale D.",
    title = "{Primordial Nucleosynthesis: from precision cosmology to fundamental physics}",
    eprint = "0809.0631",
    archivePrefix = "arXiv",
    primaryClass = "astro-ph",
    reportNumber = "DSF-20-2008, FERMILAB-PUB-08-216-A, IFIC-08-37",
    doi = "10.1016/j.physrep.2009.02.002",
    journal = "Phys. Rept.",
    volume = "472",
    pages = "1--76",
    year = "2009"
}

@article{Jedamzik:2009uy,
    author = "Jedamzik, Karsten and Pospelov, Maxim",
    title = "{Big Bang Nucleosynthesis and Particle Dark Matter}",
    eprint = "0906.2087",
    archivePrefix = "arXiv",
    primaryClass = "hep-ph",
    doi = "10.1088/1367-2630/11/10/105028",
    journal = "New J. Phys.",
    volume = "11",
    pages = "105028",
    year = "2009"
}

@article{Pospelov:2010hj,
    author = "Pospelov, Maxim and Pradler, Josef",
    title = "{Big Bang Nucleosynthesis as a Probe of New Physics}",
    eprint = "1011.1054",
    archivePrefix = "arXiv",
    primaryClass = "hep-ph",
    doi = "10.1146/annurev.nucl.012809.104521",
    journal = "Ann. Rev. Nucl. Part. Sci.",
    volume = "60",
    pages = "539--568",
    year = "2010"
}

@article{Lopez:1998vk,
    author = "Lopez, Robert E. and Turner, Michael S.",
    title = "{An Accurate Calculation of the Big Bang Prediction for the Abundance of Primordial Helium}",
    eprint = "astro-ph/9807279",
    archivePrefix = "arXiv",
    reportNumber = "FERMILAB-PUB-98-232-A",
    doi = "10.1103/PhysRevD.59.103502",
    journal = "Phys. Rev. D",
    volume = "59",
    pages = "103502",
    year = "1999"
}

@article{Burles:1999zt,
    author = "Burles, Scott and Nollett, Kenneth M. and Truran, James W. and Turner, Michael S.",
    title = "{Sharpening the predictions of big bang nucleosynthesis}",
    eprint = "astro-ph/9901157",
    archivePrefix = "arXiv",
    reportNumber = "FERMILAB-PUB-99-396-A",
    doi = "10.1103/PhysRevLett.82.4176",
    journal = "Phys. Rev. Lett.",
    volume = "82",
    pages = "4176--4179",
    year = "1999"
}

@article{Burles:2000zk,
    author = "Burles, Scott and Nollett, Kenneth M. and Turner, Michael S.",
    title = "{Big bang nucleosynthesis predictions for precision cosmology}",
    eprint = "astro-ph/0010171",
    archivePrefix = "arXiv",
    reportNumber = "FERMILAB-PUB-00-382-A",
    doi = "10.1086/320251",
    journal = "Astrophys. J. Lett.",
    volume = "552",
    pages = "L1--L6",
    year = "2001"
}

@article{Burles:2000ju,
    author = "Burles, S. and Nollett, K. M. and Turner, Michael S.",
    title = "{What is the BBN prediction for the baryon density and how reliable is it?}",
    eprint = "astro-ph/0008495",
    archivePrefix = "arXiv",
    reportNumber = "FERMILAB-PUB-00-239-A",
    doi = "10.1103/PhysRevD.63.063512",
    journal = "Phys. Rev. D",
    volume = "63",
    pages = "063512",
    year = "2001"
}

\newpage




\end{document}